\documentclass[aip,jap,reprint]{revtex4-2}

\usepackage{hyperref}
\usepackage{graphicx}
\usepackage{gensymb}
\usepackage{amsmath,amssymb}
\newcommand{\ket}[1]{\ensuremath{\left| #1 \right\rangle}}

\begin{document}

\title{Widefield quantum microscopy with nitrogen-vacancy centers in diamond: strengths, limitations, and prospects} 

\author{S. C. Scholten}
\affiliation{School of Physics, University of Melbourne, VIC 3010, Australia}

\author{A. J. Healey}
\affiliation{School of Physics, University of Melbourne, VIC 3010, Australia}
\affiliation{Centre for Quantum Computation and Communication Technology, School of Physics, University of Melbourne, VIC 3010, Australia}

\author{I. O. Robertson}
\affiliation{School of Physics, University of Melbourne, VIC 3010, Australia}

\author{G. J. Abrahams}
\affiliation{School of Physics, University of Melbourne, VIC 3010, Australia}

\author{D. A. Broadway}
\affiliation{Department of Physics, University of Basel, Klingelbergstrasse 82, Basel CH-4056, Switzerland}

\author{J.-P. Tetienne}
\email{jean-philippe.tetienne@rmit.edu.au}
\affiliation{School of Physics, University of Melbourne, VIC 3010, Australia}
\affiliation{Centre for Quantum Computation and Communication Technology, School of Physics, University of Melbourne, VIC 3010, Australia}
\affiliation{School of Science, RMIT University, Melbourne VIC 3000, Australia}

\begin{abstract}

A dense layer of nitrogen-vacancy (NV) centers near the surface of a diamond can be interrogated in a widefield optical microscope to produce spatially resolved maps of local quantities such as magnetic field, electric field and lattice strain, providing potentially valuable information about a sample or device placed in proximity. Since the first experimental realization of such a widefield NV microscope in 2010, the technology has seen rapid development and demonstration of applications in various areas across condensed matter physics, geoscience and biology. This Perspective analyzes the strengths and shortcomings of widefield NV microscopy in order to identify the most promising applications and guide future development. We begin with a brief review of quantum sensing with ensembles of NV centers, and the experimental implementation of widefield NV microscopy. We then compare this technology to alternative microscopy techniques commonly employed to probe magnetic materials and charge flow distributions. Current limitations in spatial resolution, measurement accuracy, magnetic sensitivity, operating conditions and ease of use, are discussed. Finally, we identify the technological advances that solve the aforementioned limitations, and argue that their implementation would result in a practical, accessible, high-throughput widefield NV microscope.
  
\end{abstract}

\maketitle 

\section{Introduction}

The nitrogen-vacancy (NV) defect in diamond is one of the most studied color centers in any semiconductor.\cite{dohertyNitrogenvacancyColourCentre2013} The NV has been extensively researched especially for its quantum sensing applications,\cite{degenQuantumSensing2017} which exploit the associated electron spin's sensitivity to local fields. NV-based sensing enables quantitative measurements of magnetic and electric fields, temperature and lattice strain.\cite{rondinMagnetometryNitrogenvacancyDefects2014, schirhaglNitrogenvacancyCentersDiamond2014,casolaProbingCondensedMatter2018} The electron spin resonances of the NV can be conveniently read out by optical means,\cite{gruberScanningConfocalOptical1997} making it possible to perform sensing with just a single center.\cite{balasubramanianNanoscaleImagingMagnetometry2008,mazeNanoscaleMagneticSensing2008} The small size of the NV center (its electron spin is confined within a volume of about 1\,nm diameter) and the ability to position an NV within a few nanometers of the diamond surface enable the probing of stray fields (magnetic or electric) produced by a proximate sample of interest with high spatial resolution.

\begin{figure*}[tb]
	\includegraphics[width=0.8\textwidth]{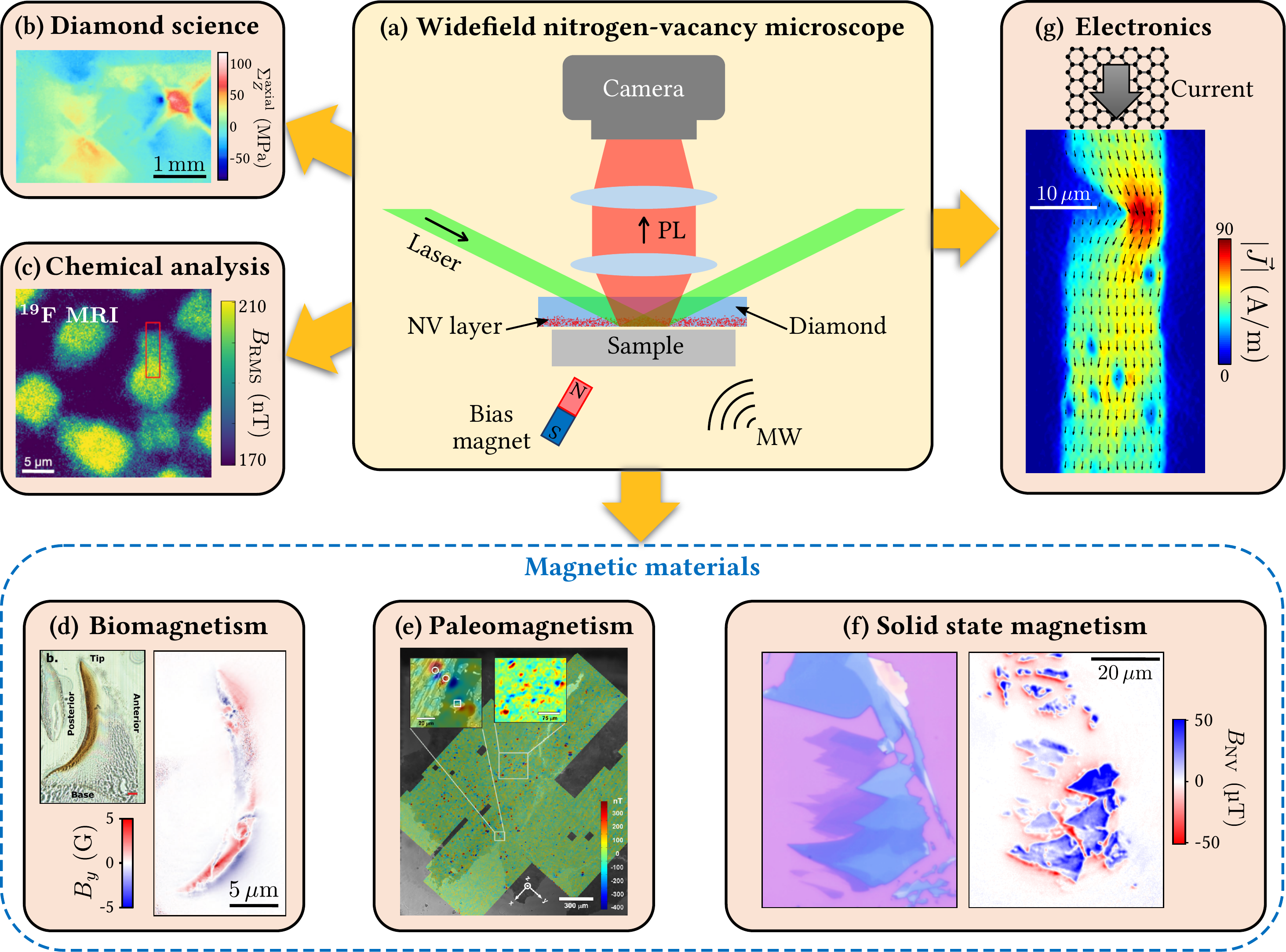}
	\caption{(a)~Schematic of a widefield NV microscope. The sample of interest is placed in proximity to a diamond chip containing a layer of NV centers near its surface. The NV layer is excited by a laser beam and the photoluminescence (PL) is imaged onto a camera. A microwave (MW) field generated by a nearby antenna is used to probe the electron spin resonances of the NV layer owing to spin-dependent PL, a technique known as optically detected magnetic resonance (ODMR). A bias magnetic field can be applied using a permanent magnet to shift the spin resonances if required. (b)-(g)~A selection of images produced by widefield NV microscopy illustrating the key capabilities and applications. (b)~Map of the total axial stress at the diamond surface, revealing growth defects. Adapted with permission from Ref.~\onlinecite{kehayiasImagingCrystalStress2019}. Copyright 2019, American Physical Society. (c)~Nuclear magnetic resonance image of CaF$_2$ islands, showing the amplitude of the $^{19}$F NMR signal. Adapted from Ref.~\onlinecite{ziemQuantitativeNanoscaleMRI2019}, under Creative Commons license CC BY 4.0 (https://creativecommons.org/licenses/by/4.0/). (d)~Stray magnetic field map of a slice of a sea mollusk's tooth (right) and corresponding optical micrograph (left). Adapted with permission from Ref.~\onlinecite{mccoeyQuantumMagneticImaging2020}. Copyright 2020, Wiley-VCH GmbH. (e)~Stray magnetic field map of a slice of a meteorite. Reproduced with permission from Ref.~\onlinecite{glennMicrometerscaleMagneticImaging2017}. Copyright 2017, American Geophysical Union. (f)~Stray magnetic field map of van der Waals VI$_3$ crystals (right) and corresponding optical micrograph (left). Adapted with permission from Ref.~\onlinecite{broadwayImagingDomainReversal2020}. Copyright 2020, Wiley-VCH GmbH. (g)~Map of the current density in a graphene ribbon reconstructed from the measured stray magnetic field. Adapted from Ref.~\onlinecite{tetienneQuantumImagingCurrent2017}, under Creative Commons license CC BY-NC 4.0 (https://creativecommons.org/licenses/by-nc/4.0/).  
	}  
	\label{Fig_intro}
\end{figure*}

Spatial mapping of these stray fields can be achieved either by scanning a single NV relative to the sample (here denoted ``scanning''),\cite{balasubramanianNanoscaleImagingMagnetometry2008,maletinskyRobustScanningDiamond2012,rondinNanoscaleMagneticField2012,tetienneNanoscaleImagingControl2014} or via a fixed dense layer of NV centers imaged onto a camera (here denoted ``widefield'').\cite{steinertHighSensitivityMagnetic2010,phamMagneticFieldImaging2011,chipauxMagneticImagingEnsemble2015,simpsonMagnetoopticalImagingThin2016} The scanning approach is technically complex but offers the highest spatial resolution, down to a few tens of nanometers.\cite{thielQuantitativeNanoscaleVortex2016} The widefield approach has a spatial resolution of 400\,nm at best but is simpler to implement due to the absence of moving parts, making it attractive for some applications. This Perspective focuses on the widefield approach to NV-based imaging, sometimes referred to as quantum diamond microscopy.\cite{glennSinglecellMagneticImaging2015,levinePrinciplesTechniquesQuantum2019} The key elements that make up a widefield NV microscope are depicted in Fig.~\ref{Fig_intro}(a). We will also discuss implementations that use nanodiamond films instead of a bulk diamond,\cite{uchiyamaOperandoAnalysisElectron2019,foyWideFieldMagneticField2020} as well as those that employ raster scanning rather than parallel imaging with a camera.\cite{maertzVectorMagneticField2010} Common to these NV microscopy variants is their optically limited spatial resolution, a key consideration when comparing with other microscopy techniques.   

Widefield imaging of dense NV layers to map stray fields was initially proposed in 2008.\cite{taylorHighsensitivityDiamondMagnetometer2008} The first experimental realization of a widefield NV microscope came in 2010, when mapping of the vector magnetic field generated by current-carrying microwires was demonstrated.\cite{steinertHighSensitivityMagnetic2010} Since then, there has been a focus on extending the capabilities and modalities of widefield NV microscopy and exploring its potential for applications in various scientific disciplines. Beside the measurement of static (DC) magnetic fields, NV microscopy has demonstrated utility in the detection of fluctuating magnetic fields at MHz and GHz frequencies, enabling electron paramagnetic resonance (EPR) and nuclear magnetic resonance (NMR) spectroscopy of liquid and solid samples.\cite{steinertMagneticSpinImaging2013,devienceNanoscaleNMRSpectroscopy2015,simpsonElectronParamagneticResonance2017,ziemQuantitativeNanoscaleMRI2019,mizunoSimultaneousWidefieldImaging2020} Key advances include the reconstruction of two-dimensional (2D) current\cite{nowodzinskiNitrogenVacancyCentersDiamond2015,tetienneQuantumImagingCurrent2017} and magnetization\cite{broadwayImagingDomainReversal2020} distributions; mapping of electric fields\cite{broadwaySpatialMappingBand2018} and strain\cite{trusheimWidefieldStrainImaging2016,broadwayMicroscopicImagingStress2019,kehayiasImagingCrystalStress2019} within the diamond; and operation at cryogenic temperatures.\cite{schlusselWideFieldImagingSuperconductor2018,lillieLaserModulationSuperconductivity2020} 

The main applications of widefield NV microscopy considered so far are summarized in Figs.~\ref{Fig_intro}(b)-\ref{Fig_intro}(g), with one example result shown to illustrate each category. Widefield NV microscopes have been employed for diamond science [Fig.~\ref{Fig_intro}(b)] for instance to characterize surface-induced magnetic noise,\cite{lillieMagneticNoiseUltrathin2018} internal electric fields,\cite{broadwaySpatialMappingBand2018} and growth-induced strain.\cite{kehayiasImagingCrystalStress2019} Electronic devices formed at the diamond surface were also investigated using electric-field mapping, providing insights into electrostatic effects during operation.\cite{broadwaySpatialMappingBand2018,lillieImagingGrapheneFieldEffect2019,lewInvestigationChargeCarrier2020} Detection and spectroscopy of fluctuating magnetic fields produced by unpaired electron or nuclear spins has been proposed as a potentially useful \textit{in situ} chemical analysis tool within a widefield NV microscope\cite{steinertMagneticSpinImaging2013,devienceNanoscaleNMRSpectroscopy2015,simpsonElectronParamagneticResonance2017,ziemQuantitativeNanoscaleMRI2019,priceWidefieldSpatiotemporalMapping2020} [Fig.~\ref{Fig_intro}(c)]. Characterizing magnetic materials by mapping their associated stray magnetic field is by far the most common use of widefield NV microscopy, with applications in biomagnetism\cite{lesageOpticalMagneticImaging2013,glennSinglecellMagneticImaging2015,davisMappingMicroscaleOrigins2018,fescenkoDiamondMagneticMicroscopy2019,mccoeyQuantumMagneticImaging2020} [Fig.~\ref{Fig_intro}(d)], paleomagnetism and rock magnetism\cite{fuSolarNebulaMagnetic2014,glennMicrometerscaleMagneticImaging2017,farchiQuantitativeVectorialMagnetic2017,fuHighSensitivityMomentMagnetometry2020,fuGaplessSpinWave2021} [Fig. \ref{Fig_intro}(e)], and solid-state magnetism [Fig.~\ref{Fig_intro}(f)]. The latter topic has seen widefield NV microscopes used to investigate a variety of artificial magnetic structures from nano- and micro-particles\cite{gouldRoomtemperatureDetectionSingle2014,tetienneProximityInducedArtefactsMagnetic2018,mccoeyRapidHighResolutionMagnetic2019} to thin-film structures,\cite{maertzVectorMagneticField2010,torailleOpticalMagnetometrySingle2018,broadwayImagingDomainReversal2020,lenzImagingTopologicalSpin2021,meirzadaLongTimeScaleMagnetizationOrdering2021,navaantonioMagneticImagingStatistical2021} as well as to characterize spin excitations in these structures\cite{bertelliMagneticResonanceImaging2020} and phase transitions at high pressure.\cite{lesikMagneticMeasurementsMicrometersized2019,hsiehImagingStressMagnetism2019,torailleCombinedSynchrotronXray2020} Finally, widefield NV microscopes have been used to probe electric currents via their associated \O rsted  magnetic field.\cite{nowodzinskiNitrogenVacancyCentersDiamond2015,tetienneQuantumImagingCurrent2017,tetienneApparentDelocalizationCurrent2019,turnerMagneticFieldFingerprinting2020} This capability was recently used to observe the viscous Dirac fluid nature of current in graphene,\cite{kuImagingViscousFlow2020} and to image photocurrent vortices in MoS$_2$.\cite{zhouSpatiotemporalMappingPhotocurrent2020} Applications in electronics also include probing currents in the microwave regime\cite{shaoDiamondRadioReceiver2016,horsleyMicrowaveDeviceCharacterization2018,marianiSystemRemoteControl2020} and temperature mapping of electronic circuits.\cite{andrichMicroscaleResolutionThermalMapping2018,uchiyamaOperandoAnalysisElectron2019,foyWideFieldMagneticField2020,chenSimultaneousImagingMagnetic2021}

The purpose of this Perspective is to analyze some of these applications in more detail and, considering the current state of the art of widefield NV microscopy, identify the strengths and shortcomings compared to other microscopy techniques typically used. The key features of widefield NV microscopy that are often put forward are its high magnetic sensitivity, its versatility and simplicity, its high spatial resolution, and its ability to accurately measure a sample's magnetic properties. Our discussions will therefore aim to put these perceived advantages within a broader context. The article is organized as follows. We will begin with a brief review of the key techniques utilized in widefield NV microscopy experiments (Sec.~\ref{sec:principles}), which will inform  the subsequent discussions. More detailed treatments can be found in the literature regarding magnetic sensing using ensembles of NV centers\cite{barrySensitivityOptimizationNVdiamond2020} and on the principles and techniques of widefield NV microscopy.\cite{levinePrinciplesTechniquesQuantum2019} Focusing on magnetic field imaging, we will then discuss the advantages and limitations of widefield NV microscopy compared to standard techniques that can be used to gain similar information (Sec.~\ref{sec:comparison}). In Sec.~\ref{sec:limitations}, we will discuss in more detail the current limits to spatial resolution, measurement accuracy, magnetic sensitivity, operating conditions and ease of use, and discuss future technological developments to improve those aspects.

\section{\label{sec:principles}Principles of widefield NV microscopy} 

Here we briefly outline some key principles of widefield NV microscopy that will be relevant to the subsequent discussions. First, we provide a synthetic review of sensing with ensembles of NV centers, highlighting how the effects of magnetic field, electric field, strain and temperature can be distinguished. We then describe the different implementations of a widefield NV microscopy experiment, specifically in the interfacing of the diamond sensor with the sample under study.

\subsection{Sensing with ensembles of NV centers}

\begin{figure*}[tb]
	\includegraphics[width=0.9\textwidth]{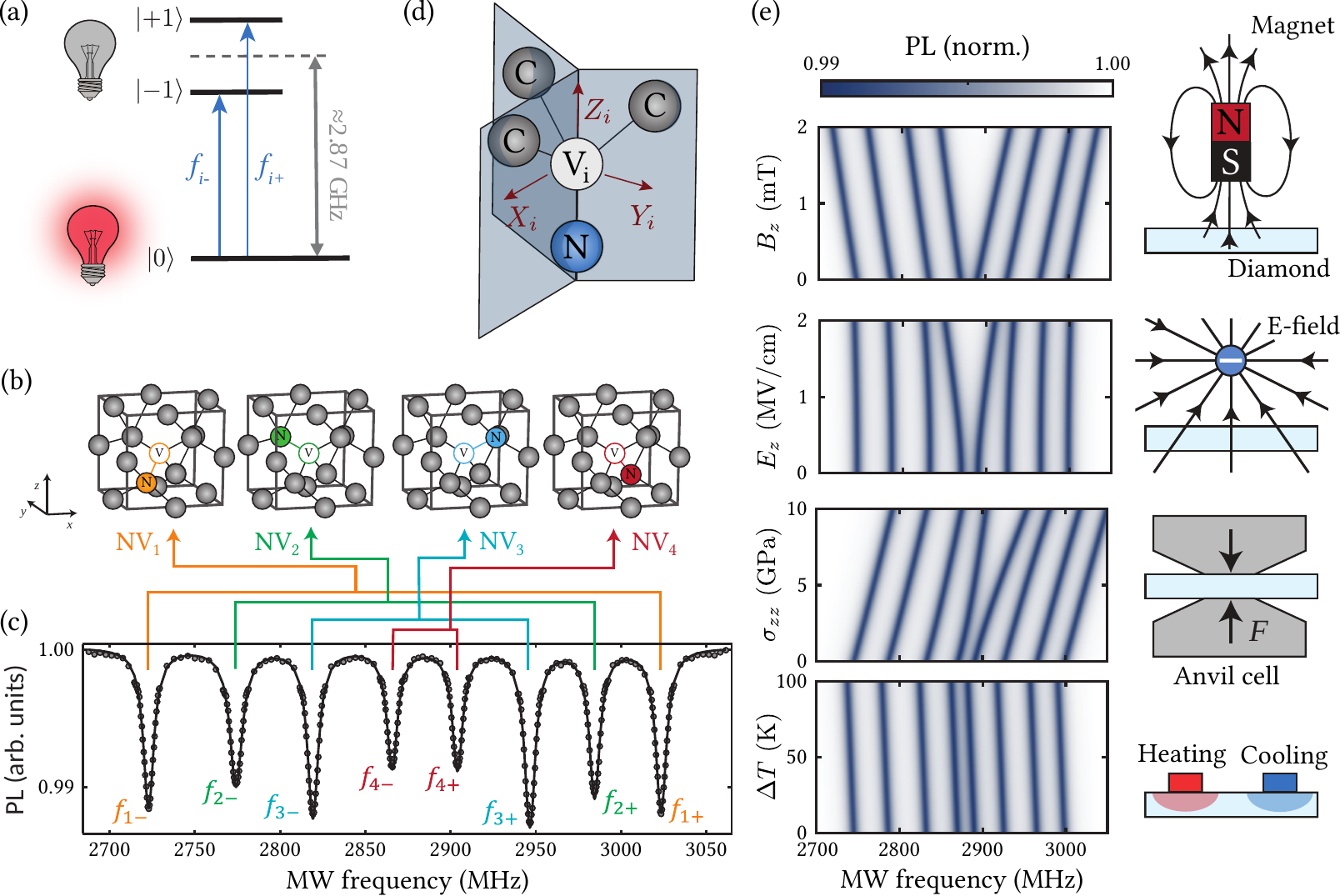}
	\caption{(a)~Energy level diagram of the NV center's ground state. The two electron spin transitions with frequencies $f_{i\pm}$ differ in the presence of nonzero field or strain. (b)~Representation of an NV center in the diamond crystal lattice for the four possible orientations (labeled NV$_i$ with $i=1,2,3,4$), and definition of the lab-frame coordinate system ($xyz$). (c)~Optically detected magnetic resonance (ODMR) spectrum of an ensemble of NV centers in a single-crystal diamond recorded under a bias magnetic field $\vec{B}_0\approx(-2.5,-1.2,4.2)$\,mT (expressed in the $xyz$ frame). The direction of $\vec{B}_0$ was chosen in order to allow each NV orientation to be resolved. (d)~Defect structure of the NV and its native coordinate system ($X_i,Y_i,Z_i$). (e)~Evolution of the eight spin resonances as a function of an external stimulus, given a fixed initial bias magnetic field as in (c). Each graph corresponds to, from top to bottom: a magnetic field $B_z$, an electric field $E_z$, an axial stress $\sigma_{zz}$, and a temperature change $\Delta T$.
	}  
	\label{Fig_ODMR}
\end{figure*}

Sensing using the NV center in diamond relies on the fine structure of its electronic ground state; a spin triplet ($S=1$) with a zero-field splitting of $D\approx2.87$\,GHz between the $\ket{0}$ state and the $\ket{\pm}$ states [Fig.~\ref{Fig_ODMR}(a)]. Owing to spin-dependent photoluminescence (PL) combined with an optical pumping effect,\cite{dohertyNitrogenvacancyColourCentre2013} it is possible to probe the two spin transitions $\ket{0}\rightarrow\ket{\pm}$ via optically detected magnetic resonance (ODMR) and measure the corresponding frequencies $f_{i\pm}$. The presence of a nonzero magnetic field, electric field, lattice strain, or a change in temperature, shift the energy levels and consequently change the transition frequencies $f_{i\pm}$. The precise measurement of these changes, whether static (constant shift of $f_{i\pm}$) or dynamic (fluctuations of $f_{i\pm}$), underpins most NV sensing experiments. 

In a widefield NV microscope, one often exploits the fact that a single-crystal diamond can have NV centers oriented along four possible crystallographic directions $\langle 111\rangle$, represented in Fig.~\ref{Fig_ODMR}(b). By applying a bias magnetic field $\vec{B}_0$, it is possible to isolate any desired orientation family, and even to measure them all at once. This is illustrated in Fig.~\ref{Fig_ODMR}(c), which shows an ODMR spectrum obtained in a field $B_0\approx6$\,mT oriented such that all four orientation families (eight resonances in total) can be resolved. Multiple resonance frequencies are most conveniently obtained by recording a complete ODMR spectrum as in Fig.~\ref{Fig_ODMR}(c), but multi-channel Ramsey interferometry techniques can also be employed.\cite{levinePrinciplesTechniquesQuantum2019,schlossSimultaneousBroadbandVector2018,hartDiamondMagneticMicroscopy2021} 

For a given NV orientation, the frequencies $f_{i\pm}$ may be modeled by a simplified spin Hamiltonian (in units of frequency, Hz),\cite{dohertyTheoryGroundstateSpin2012,udvarhelyiSpinstrainInteractionNitrogenvacancy2018,barfussSpinstressSpinstrainCoupling2019}
\begin{align} \label{eq:Ham}
H_i &= (D + \mathcal{M}_{Z_i})S_{Z_i}^2 + \gamma_{\rm NV}\vec{B}\cdot\vec{S}_i  \\
&\quad - \mathcal{M}_{X_i} ( S_{X_i}^2 - S_{Y_i}^2)  + \mathcal{M}_{Y_i}( S_{X_i}S_{Y_i} + S_{Y_i}S_{X_i}),
\end{align}
where $\gamma_{\rm NV}=28.035(3)$\,GHz T$^{-1}$ is the NV gyromagnetic ratio,\cite{dohertyNitrogenvacancyColourCentre2013} $\vec{S}_i=(S_{X_i},S_{Y_i},S_{Z_i})$ are the spin-1 operators, $\vec{B}$ is the applied magnetic field, and $\vec{\cal M}_i=({\cal M}_{X_i},{\cal M}_{Y_i},{\cal M}_{Z_i})$ is a vector that captures the effects of electric field and lattice strain, expressed in the coordinate system of the NV defect structure, $(X_i,Y_i,Z_i)$ [Fig.~\ref{Fig_ODMR}(d)], with the index $i=1,2,3,4$ denoting the NV orientation with respect to the lab frame [Fig.~\ref{Fig_ODMR}(b)]. We note that the parameter $D$ depends on temperature,\cite{acostaTemperatureDependenceNitrogenVacancy2010,dohertyTemperatureShiftsResonances2014} and so Eq.~(\ref{eq:Ham}) describes the combined effects of magnetic field, electric field, strain and temperature.

\subsubsection{Magnetic field sensing} \label{sec:Bsensing}

For magnetic field measurements, the effect of $\vec{\cal M}_i$ is typically neglected. In this case, there are four unknown quantities $(D,B_x,B_y,B_z)$, which can be uniquely retrieved by measuring at least two NV orientation families (giving four frequencies in total). In practice, one generally measures all four families to maximise the accuracy and sensitivity of the measurement of the vector field $\vec{B}$, either using a single 8-frequency measurement as in Fig.~\ref{Fig_ODMR}(c), or using sequential 2-frequency measurements under different bias field conditions.\cite{levinePrinciplesTechniquesQuantum2019} The field $\vec{B}$ is the sum of the known bias field and the stray field emanating from the sample under study, $\vec{B}=\vec{B}_0+\vec{B}_{\rm sample}$, and so the sample field can be deduced by simple subtraction. The magnetic field maps shown in Figs.~\ref{Fig_intro}(d) and~\ref{Fig_intro}(e) were obtained using this vector magnetometry method, where only one vector component is displayed for conciseness.    

For some applications, it is preferable or sufficient to measure only one NV orientation family. In this case, the bias field is aligned along the chosen orientation, and the transition frequencies are given by
\begin{align} \label{eq:aligned}
f_{i\pm} &= D\pm\gamma_{\rm NV}B_{Z_i}
\end{align}
where $B_{Z_i}$ is the projection of $\vec{B}$ along the symmetry axis ($Z_i$) of this orientation. When possible, the two frequencies $f_{i\pm}$ are measured, so that the field projection can be extracted independent of temperature or strain contributions,    
\begin{align} \label{eq:BZ}
B_{Z_i} = \frac{f_{i+}-f_{i-}}{2\gamma_{\rm NV}}.
\end{align}
The magnetic field map in Fig.~\ref{Fig_intro}(f) was obtained in this way. 

\subsubsection{Sensing of other quantities}

We now briefly discuss how strain, electric field, and temperature can be extracted from ODMR measurements. For a given NV orientation, the vector $\vec{\cal M}_i$, which has units of frequency, is given by\cite{broadwayMicroscopicImagingStress2019}
\begin{align} \label{eq:M}
\mathcal{M}_{X_i} &= k_\perp E_{X_i} + b\Sigma_X^{\rm axial} + c\Sigma_{X_i}^{{\rm shear}} \nonumber \\
\mathcal{M}_{Y_i} &= k_\perp E_{Y_i} + \sqrt{3}b\Sigma_Y^{\rm axial} + \sqrt{3}c\Sigma_{Y_i}^{{\rm shear}} \\
\mathcal{M}_{Z_i} &= a_1\Sigma_Z^{\rm axial}  + 2a_2\Sigma_{Z_i}^{{\rm shear}} \nonumber
\end{align}
where $\vec{E}=(E_{X_i},E_{Y_i},E_{Z_i})$ is the electric field vector projected onto the NV frame, $k_\perp=17(3)$\,Hz\,cm\,V$^{-1}$ is the transverse electric susceptibility parameter,\cite{doldeElectricfieldSensingUsing2011} $a_1 = 4.86(2)$, $a_2 = -3.7(2)$, $2b = -2.3(3)$, and $2c = 3.5(3)$ in units of MHz GPa$^{-1}$ are the stress susceptibility parameters.\cite{barsonNanomechanicalSensingUsing2017,barfussSpinstressSpinstrainCoupling2019} We also introduced the quantities   
\begin{align} \label{eq:axial}
\Sigma_X^{\rm axial} &= -\sigma_{xx}-\sigma_{yy}+2\sigma_{zz} \nonumber \\
\Sigma_Y^{\rm axial} &= \sigma_{xx} - \sigma_{yy} \\
\Sigma_Z^{\rm axial} &= \sigma_{xx}+\sigma_{yy}+\sigma_{zz}, \nonumber
\end{align}
which capture the contribution of the axial stress components and are independent of the NV orientation, and
\begin{align} \label{eq:shear}
\Sigma_{X_i}^{{\rm shear}} &= 2\alpha_i \sigma_{xy}+\beta_i\sigma_{xz}+\alpha_i\beta_i\sigma_{yz} \nonumber \\
\Sigma_{Y_i}^{{\rm shear}} &= \beta_i\sigma_{xz}-\alpha_i\beta_i \sigma_{yz} \\
\Sigma_{Z_i}^{{\rm shear}} &= \alpha_i\sigma_{xy} -\beta_i \sigma_{xz} -\alpha_i\beta_i \sigma_{yz}, \nonumber
\end{align}
which describe the effect of the shear stress components. Here the stress tensor $\overleftrightarrow\sigma$ was expressed using its six independent lab-frame components ($\sigma_{xx},\sigma_{yy},\sigma_{zz},\sigma_{xy},\sigma_{xz},\sigma_{yz}$). In Eq.~(\ref{eq:shear}), the functions $\alpha_i$ and $\beta_i$ evaluate to $\pm1$ depending on the NV orientation. Namely, using the NV orientations $i=1,2,3,4$ as defined in Fig.~\ref{Fig_ODMR}(b), we have $(\alpha_1,\alpha_2,\alpha_3,\alpha_4)=(+1,-1,-1,+1)$ and $(\beta_1,\beta_2,\beta_3,\beta_4)=(-1,+1,-1,+1)$. In Eq.~(\ref{eq:M}), we did not include a longitudinal electric field term ($E_{Z_i}$) because the effect of this term averages out in the case of an ensemble of NV centers (due to the two possible axial orientations for each NV family\cite{dohertyMeasuringDefectStructure2014,broadwaySpatialMappingBand2018}).  

In general, with a total of 13 unknown quantities ($D$, magnetic field vector, electric field vector, stress tensor), it is not possible to infer those parameters simply by applying a known bias magnetic field. However, under appropriate assumptions and approximations, it is possible to extract several of these unknown quantities simultaneously from an 8-frequency measurement as in Fig.~\ref{Fig_ODMR}(c). For instance, if strain is neglected ($\overleftrightarrow\sigma=0$), then all 7 remaining unknowns can in principle be inferred.\cite{broadwaySpatialMappingBand2018} Conversely, if  the electric field is neglected ($\vec{E}=0$) and the vector magnetic field $\vec{B}$ is known or sufficiently small, then all 3 shear stress components ($\sigma_{xy},\sigma_{xz},\sigma_{yz}$) as well as the sum $D+a_1\Sigma_Z^{\rm axial}$ can be obtained,\cite{kehayiasImagingCrystalStress2019,broadwayMicroscopicImagingStress2019} and in some conditions the individual axial stress components can also be inferred.\cite{broadwayMicroscopicImagingStress2019} Figure~\ref{Fig_intro}(a) shows a map of the total axial stress component ($\Sigma_Z^{\rm axial}$) in the NV layer at the surface of a diamond, obtained using these methods. 

The various effects are summarized in Fig.~\ref{Fig_ODMR}(e), which plots the evolution of the 8 ODMR transition frequencies under various stimuli, given a constant bias magnetic field. An additional magnetic field (here $B_z$) changes the splitting of the resonances ($f_{i\pm}$) by a similar amount for the different NV families. In contrast, an electric field is characterized by a stronger splitting effect on the NV families that are least split to begin with (where the NV axis is orientated more perpendicular to the bias field). Strain splits the resonances in a similar fashion to electric field but is always accompanied by a common-mode shift of larger magnitude than (or at least comparable to) the splitting. Finally, a change in temperature has a distinct signature because it shifts all the resonances by the same amount (by changing the shared parameter $D$), without any change to the splittings. These qualitative observations illustrate how it is possible to separate different effects and perform multimodal sensing with NV ensembles.     

\subsubsection{Applications of multimodal sensing}

The ability to perform multimodal imaging in a widefield NV microscope can be useful in diamond science. For instance, all-diamond electronic devices based on the 2D hole gas formed near the diamond surface have been studied using simultaneous magnetic field and electric field mapping, providing insights into the relationship between charge transport and electrostatic effects.\cite{broadwaySpatialMappingBand2018} Mapping surface-induced electric fields and strain is also a useful tool to help optimize diamond samples for quantum sensing applications including widefield NV microscopy.\cite{broadwaySpatialMappingBand2018,broadwayMicroscopicImagingStress2019,kehayiasImagingCrystalStress2019} For studies of samples and devices external to the diamond, electric field sensing appears more challenging because of screening by the diamond surface,\cite{obergSolutionElectricField2020} although this remains an open research direction.\cite{bianNanoscaleElectricfieldImaging2021,barsonNanoscaleVectorElectric2021}

Measuring temperature alongside magnetic fields in a widefield NV microscope is also of potential interest for electronics applications, enabling the \textit{in situ} monitoring of Joule heating while imaging the current density in an operating device.\cite{andrichMicroscaleResolutionThermalMapping2018,uchiyamaOperandoAnalysisElectron2019,foyWideFieldMagneticField2020,chenSimultaneousImagingMagnetic2021} However, a single-crystal diamond in contact with the device of interest thermalizes rapidly and acts as a heat sink, and so these applications are often tackled using a modified version of a widefield NV microscope whereby a layer of nanodiamonds that are thermally decoupled from each other is formed onto the electronic device.\cite{andrichMicroscaleResolutionThermalMapping2018,uchiyamaOperandoAnalysisElectron2019,foyWideFieldMagneticField2020} The temperature response of the $D$ parameter also becomes vanishingly small at cryogenic temperatures (below $100$\,K),\cite{dohertyTemperatureShiftsResonances2014} which prevents \textit{in situ} temperature monitoring in cryogenic widefield NV microscopy experiments.\cite{lillieLaserModulationSuperconductivity2020}

In contrast, magnetic field sensing with NV centers performs similarly at room and cryogenic temperatures, and does not suffer from screening effects because most materials including the diamond are magnetically transparent. As a result, the vast majority of widefield NV microscopy applications are concerned with magnetic field sensing only, which will be our focus for the rest of this article.

\subsubsection{\label{subsec:ac}Sensing of time-varying signals} 

NV-based magnetometry is not restricted to the measurement of DC signals, and techniques have been developed to allow the measurement of alternating (AC) and randomly fluctuating signals in the kHz to GHz regime. These techniques are generally compatible with widefield NV microscopy.\cite{levinePrinciplesTechniquesQuantum2019}

The sensing of up to MHz signals can be facilitated by the use of microwave pulse sequences such as XY8, where the pulse spacing allows the NV ensemble to ``lock in" to a target signal and decouple from other sources of noise.\cite{delangeSingleSpinMagnetometryMultipulse2011,staudacherNuclearMagneticResonance2013} The sensitivity of the measurement to both the frequency and amplitude of the signal is set by the length of the sequence used, in practice limited by the NV coherence time $T_2$. An upper limit to the frequencies accessible using dynamical decoupling sequences is imposed by the width of the microwave pulses used; the frequency of the signal to be measured must be much lower than the Rabi driving frequency which can be up to order 100\,MHz in widefield experiments. Dynamical decoupling is most commonly used to perform NMR spectroscopy, where the statistical polarization of a nuclear spin ensemble is measured using shallow NVs.\cite{staudacherNuclearMagneticResonance2013} This technique has been adapted to widefield NV microscopy, enabling spatially resolved NMR with a sub-micron resolution.\cite{devienceNanoscaleNMRSpectroscopy2015,ziemQuantitativeNanoscaleMRI2019}  

Where the NV transition frequency $f_{i\pm }$ can be brought into resonance with a target field, simpler schemes can be used and GHz frequencies accessed. The transition between NV spin states can be driven coherently and the strength of the target field inferred by the Rabi oscillation frequency, as recently demonstrated to characterize microwave devices\cite{horsleyMicrowaveDeviceCharacterization2018} and to image coherent spin waves.\cite{bertelliMagneticResonanceImaging2020} Resonant magnetic noise can also accelerate decay to a thermal mixture from an initially polarized $\ket{0}$ state, enabling a microwave-free measurement scheme dubbed $T_1$ relaxometry.\cite{steinertMagneticSpinImaging2013,tetienneSpinRelaxometrySingle2013,kaufmannDetectionAtomicSpin2013} This technique has been used to image paramagnetic spins\cite{steinertMagneticSpinImaging2013,simpsonElectronParamagneticResonance2017} and antiferromagnetic domain walls,\cite{fincoImagingNoncollinearAntiferromagnetic2021} and again requires shallow NVs due to the rapidly decaying nature of these signals. In both cases, the spectral selectivity is given by the ODMR linewidth and can also manifest as a reduction in ODMR contrast.\cite{lillieMagneticNoiseUltrathin2018} 

Complicating the widefield operation of $T_1$-based protocols is the possible conflation of variation in NV charge state across the field of view with a target field. In particular, the charge dynamics will typically be highly correlated with the laser spot profile, which is usually not uniform over the full field of view. Further, the NV charge state is particularly unstable in the required near-surface regime.\cite{bluvsteinExtendingQuantumCoherence2019} Appropriate normalization and experimental design can overcome these issues, leaving the prospects for widefield $T_1$ relaxometry strong. 

Finally, NVs are also capable of probing repeatable transient phenomena in a stroboscopic manner with a temporal resolution of about 100\,ns, for typical Rabi driving frequency of order 10\,MHz. This method has recently been used to map in space and time the rise of a photocurrent vortex following a light pulse.\cite{zhouSpatiotemporalMappingPhotocurrent2020} 

\subsection{\label{sec:implementation}Implementation of widefield NV microscopy}

\begin{figure*}[tb]
	\includegraphics[width=0.99\textwidth]{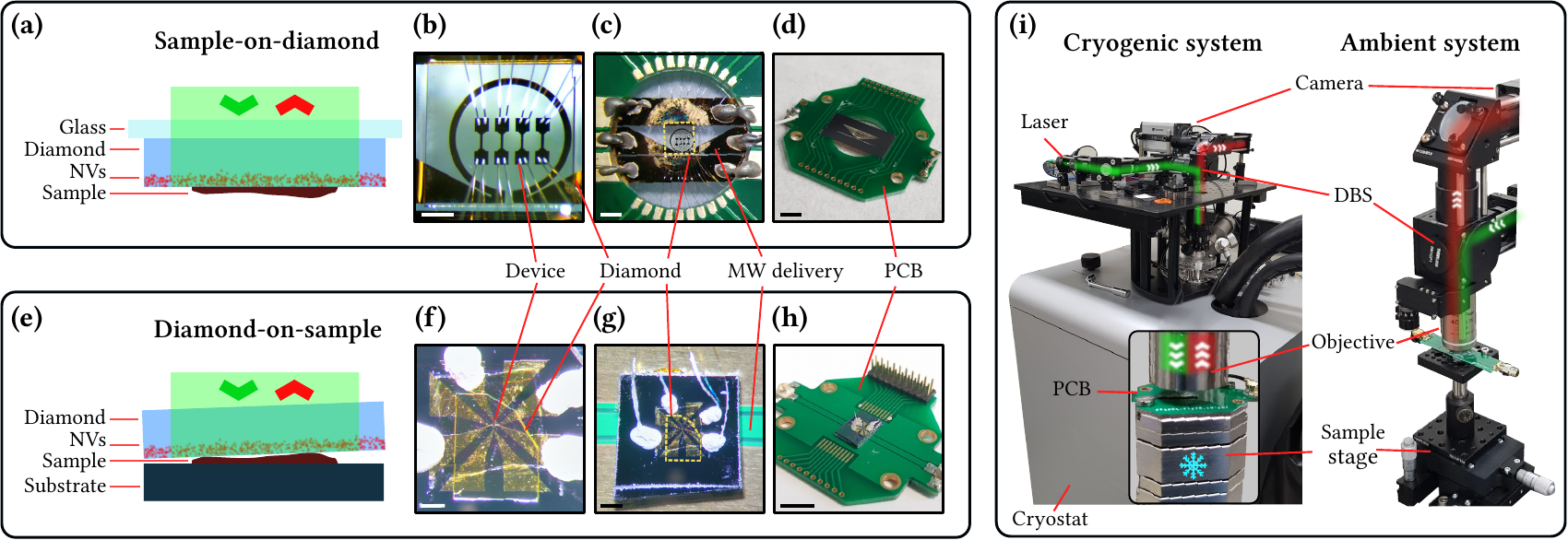}
	\caption{Implementation of a widefield NV microscope. (a)~Schematic depicting the sample-on-diamond interfacing method. (b)~Photograph of niobium devices fabricated directly on the diamond (transparent square) which is glued to a glass coverslip patterned with a gold omega-shaped microwave (MW) resonator. Scalebar:~500\,$\mu$m. (c)~Same sample/coverslip as (b)~mounted on a PCB with a yellow dotted line highlighting the lateral extent of the diamond. The PCB is connected to the niobium devices with aluminum wirebonds and to the resonator on the glass coverslip via silver epoxy. Scalebar:~2\,mm. (d)~The PCB in (c)~is upside-down emphasizing the imaging conditions. Scalebar:~5\,mm. (e)~Schematic depicting the diamond-on-sample interfacing method. (f)~Photograph of electrical devices with a diamond (transparent polygon) glued on top. Scalebar:~500\,$\mu$m. (g)~Same sample as (f)~glued to a PCB. The PCB has an integrated stripline waveguide for microwave delivery (green horizontal region), and is connected to the device via aluminum wirebonds. Scalebar:~1\,mm. (h)~Similar PCB and sample to (g)~emphasizing the upright imaging conditions. Scalebar:~5\,mm. (i)~Photographs of two upright widefield NV microscopes: cryogenic (left) and ambient (right). The optics for both setups are similar, addressing the NVs with a 532\,nm laser and objective which is simultaneously used for readout. The red PL is separated with a DBS before being imaged with a sCMOS camera. (Inset) the sample mount contained within the cryostat chamber with a piezo actuated 3-axis stage, cold temperature objective and PCB (note similarity to (d) and (h)). Not shown is the superconducting vector magnet in the cryostat system or the permanent magnet utilized in the ambient system for biasing. PCB: printed circuit board, DBS: dichroic beam-splitter, sCMOS: scientific complementary metal-oxide-semiconductor.
	}  
	\label{Fig_implem}
\end{figure*}

The key components of a widefield NV microscope are depicted in Fig.~\ref{Fig_intro}(a). However, the details of its experimental implementation often depends on the specific application. Here we review a few key technical aspects, namely the diamond-sample interfacing, the optical microscope setup, and diamond sensor fabrication. 

\subsubsection{\label{subsec:interfacing}Sensor-sample interfacing} 

An important design consideration in the planning of a widefield NV microscopy experiment is the interfacing of the sample or device of interest with the diamond sensor. Two main approaches are typically employed, depicted in Figs.~\ref{Fig_implem}(a)-\ref{Fig_implem}(h). The first approach involves direct deposition of the sample onto the diamond surface [Fig.~\ref{Fig_implem}(a); here denoted ``sample-on-diamond'']. Alternatively, the diamond may be mechanically brought near or in contact with the same [Fig.~\ref{Fig_implem}(e); here denoted ``diamond-on-sample'']. 

The diamond-on-sample approach is simpler in that the sample does not require a special preparation for widefield NV microscopy,\cite{simpsonMagnetoopticalImagingThin2016} it can even be a complex device including some electrical contacts, or a large slice of an animal or a rock.\cite{glennMicrometerscaleMagneticImaging2017} The diamond can be simply dropped on the sample, or be brought in proximity using a positioning stage. Figures~\ref{Fig_implem}(f)-\ref{Fig_implem}(h) illustrate this approach with photos of a diamond (lateral size $\sim$\,1\,mm$^2$) dropped NV face down on an electronic device fabricated on a Si substrate. One drawback of this approach is that the standoff distance between the NV layer and the sample is not well controlled, especially if the sample is not flat, which may limit the spatial resolution (see Sec.~\ref{subsec:spatial_res}) and complicate data analysis. The minimum standoff depends on the flatness and roughness of the diamond and sample, but is often on the order of a few microns.\cite{glennMicrometerscaleMagneticImaging2017,bertelliMagneticResonanceImaging2020} Figure~\ref{Fig_intro}(e) shows a magnetic field image obtained using this approach, where the diamond was placed on a thin section of a meteorite.\cite{glennMicrometerscaleMagneticImaging2017} 

In contrast, the sample-on-diamond approach often yields a well-defined geometry if the sample is allowed to conform to the diamond surface, as in the case of particles or deposited thin films. In this case, the standoff distance can be as small as 10\,nm, given by the depth of the NV layer.\cite{lesageOpticalMagneticImaging2013,tetienneQuantumImagingCurrent2017} A very small standoff can be advantageous for sensing of minute fluctuating magnetic fields,\cite{devienceNanoscaleNMRSpectroscopy2015,simpsonElectronParamagneticResonance2017} although for DC imaging a standoff of order 100\,nm is sufficient to reach the optical diffraction limit in the lateral spatial resolution.\cite{levinePrinciplesTechniquesQuantum2019,healeyComparisonDifferentMethods2020} The main drawback of this approach is the difficulty of sample fabrication compared with a standard substrate, although complex electronic devices have been successfully fabricated on diamond for widefield NV microscopy purposes.\cite{lillieImagingGrapheneFieldEffect2019,kuImagingViscousFlow2020} Figures~\ref{Fig_implem}(b)-\ref{Fig_implem}(d) show example photos of electronic devices directly fabricated on a 2\,mm\,$\times$\,2\,mm diamond, which is itself mounted on a glass coverslip. The images in Figs.~\ref{Fig_intro}(c),\ref{Fig_intro}(f) and \ref{Fig_intro}(g) were all obtained using this approach.

\subsubsection{Optical microscope setup}

The widefield fluorescence microscope necessary to image the NV layer can take different forms, often dictated by the application. Photos of two different implementations of a widefield NV microscope are shown in Fig.~\ref{Fig_implem}(i), with the sample either in ambient conditions (right) or in a cryostat (left). A custom-built microscope gives the most flexibility, but widefield NV microscopy experiments based on a commercial optical microscope have also been reported.\cite{simpsonMagnetoopticalImagingThin2016,priceWidefieldSpatiotemporalMapping2020} Both upright and inverted microscope configurations can be employed, although the upright configuration [as is the case in Fig.~\ref{Fig_implem}(i)] is generally preferred with the diamond-on-sample approach (to prevent the diamond from falling). Laser excitation of the NV layer is most often achieved either by side illumination as depicted in Fig.~\ref{Fig_intro}(a), or by epifluorescence illumination (where laser and fluorescence paths are collinear and travel through the same objective lens) as is the case in Fig.~\ref{Fig_implem}(i). Considerations for the design and choice of the various optical components (including laser and camera) are discussed extensively in Ref.~\onlinecite{levinePrinciplesTechniquesQuantum2019}.

\subsubsection{\label{subsec:sensor_design}Sensor design and fabrication} 

\begin{figure}[bt]
	\includegraphics[width=0.5\textwidth]{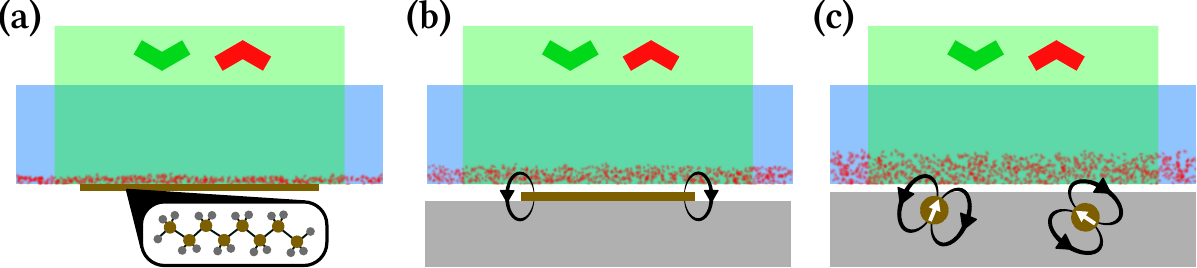}
	\caption{Sensor design. The thickness of the NV layer is chosen according to the sample under study: (a) $\sim$\,10\,nm for rapidly decaying fluctuating signals from e.g.\ nuclear spins in molecules; (b) 0.1-1\,$\mu$m for static magnetic signals from thin films; (c) $\sim$\,10\,$\mu$m for extended or buried samples.    
	}  
	\label{Fig_thickness}
\end{figure}

The central component of a widefield NV microscope is the NV sensing layer. As a widefield microscope collects fluorescence from throughout the diamond, it is important to confine the NVs in a layer of the desired thickness (typically from 10\,nm to 10\,$\mu$m), with no NVs elsewhere. The NV layer is generally formed directly underneath the diamond surface to ensure close proximity with the sample, although a capping or spacing layer can be added if necessary. The diamond as a whole should ideally be thin ($\sim$\,100\,$\mu$m) to minimize optical losses and aberrations.

The thickness of the sensing layer impacts both the spatial resolution and the magnetic sensitivity, as will be discussed later (Sec.~\ref{subsec:spatial_res} and~\ref{subsec:field_sens}). Therefore, it is usual to tailor the layer thickness to the requirements of the intended application,\cite{levinePrinciplesTechniquesQuantum2019} as illustrated in Fig.~\ref{Fig_thickness}. For example, imaging of EPR/NMR signals requires a shallow layer (approximately 10\,nm) to maximize the detection of the rapidly decaying signal,\cite{devienceNanoscaleNMRSpectroscopy2015,ziemQuantitativeNanoscaleMRI2019} micron-scale imaging of magnetic thin films prefers 0.1-1\,$\mu$m layers,\cite{broadwayImagingDomainReversal2020} and mm-scale imaging of current distributions and geological samples may employ layers 10\,$\mu$m thick or more.\cite{glennMicrometerscaleMagneticImaging2017,turnerMagneticFieldFingerprinting2020}

Several methods exist to create such NV layers. Growth of diamond layers by chemical vapor deposition (CVD) with nitrogen co-doping, followed by electron irradiation to create the vacancies, has the ability to create NV layers of virtually any thickness and NV density.\cite{ohnoEngineeringShallowSpins2012,smithColourCentreGeneration2019,achardChemicalVapourDeposition2020} However, the CVD method remains difficult and costly to perfect.

An alternative, cost-effective technique is ion implantation of commercially available high-pressure high-temperature (HPHT) grown diamond.\cite{huangDiamondNitrogenvacancyCenters2013,mccloskeyHeliumIonMicroscope2014} This method is suitable for creating 0.1-1\,$\mu$m thick NV layers and  can produce comparable sensitivity to CVD layers, although with less control over the NV density.\cite{healeyComparisonDifferentMethods2020} Here an NV layer is produced by taking advantage of the fact that although the diamonds contain nitrogen throughout (typically at high concentrations, $\sim$\,100\,ppm), the default NV conversion is extremely low, and so localizing vacancies through ion implantation is successful in producing a region that dominates the total fluorescence. The width of the vacancy production profile scales with the size of the ion implanted, so using lighter ions such as He and C can produce sub-100\,nm layers while thicker layers (up to 1\,$\mu$m) can be activated using heavier ions.

Finally, implantation of nitrogen-free diamond with nitrogen ions is an effective method for creating shallow NV layers (of order 10\,nm thick),\cite{pezzagnaCreationEfficiencyNitrogenvacancy2010,tetienneSpinPropertiesDense2018} but offers inferior sensitivity to the other methods for thicker layers.\cite{healeyComparisonDifferentMethods2020}  Further detail on diamond materials considerations and the impacts on magnetic sensitivity are reserved for Sec.~\ref{subsec:field_sens}.

Another design consideration is the crystallographic orientation of the substrate. Most common diamonds have \{100\} principal faces. These substrates are optimal for vector magnetometry as a linear microwave field polarized in $z$ (i.e.\ with a simple loop resonator) can address all NV orientations equally. For single-axis measurements it is sometimes preferable to be most sensitive to stray field oriented out-of-plane, in which case \{111\} diamonds can be used. This geometry is also convenient in the specific case of imaging ultrathin magnets with out-of-plane anisotropy at different fields along their preferred axis. Single-axis measurements also benefit from engineering preferentially oriented NV ensembles, which can be achieved through CVD growth on \{111\} and \{113\} substrates.\cite{lesikPerfectPreferentialOrientation2014,lesikPreferentialOrientationNV2015,osterkampEngineeringPreferentiallyalignedNitrogenvacancy2019} \{110\} cut diamonds can also be of use with their two in-plane NVs, for example in 2D current density imaging.\cite{broadwayImprovedCurrentDensity2020}  Further development to increase the availability of these less common cuts would improve the versatility of widefield NV microscopy.

\section{\label{sec:comparison}Comparison with other techniques} 

Having reviewed the basics of sensing with NV ensembles and the experimental implementation of widefield NV microscopy, we now move on to discuss the strengths and shortcomings of this technology and a comparison with alternative techniques that can be employed to gain similar information about a sample or device. We begin by discussing the imaging of static magnetic field distributions, which is by far the most used imaging modality of widefield NV microscopy. The imaging of time-varying magnetic signals up to GHz frequencies will then be addressed. 

\subsection{Static magnetic imaging}

Microscopic imaging of static magnetic fields is a valuable tool to study magnetic materials via the stray magnetic field they produce, as well as charge currents in electronic devices via their induced \O rsted magnetic field. Analysis can also include the study of the magnetic response of nonmagnetic materials (paramagnetism and diamagnetism, including superconductors), which may produce a measurable stray field in response to an applied magnetic field. The imaging techniques available to study magnetic materials and current distributions can be divided into two classes. The first one encompasses techniques that are sensitive to the stray magnetic field produced by a sample, as with NV microscopy. The second class of relevant imaging techniques relies on the interaction of test particles (e.g.\ photons or electrons) with the sample, in such a way that the interaction depends on the magnetization of the sample (for magnetic materials) or the current distribution. Figure~\ref{Fig_comparison} illustrates some of these techniques. 

\subsubsection{Stray-field techniques}

\begin{figure}[tb]
	\includegraphics[width=0.4\textwidth]{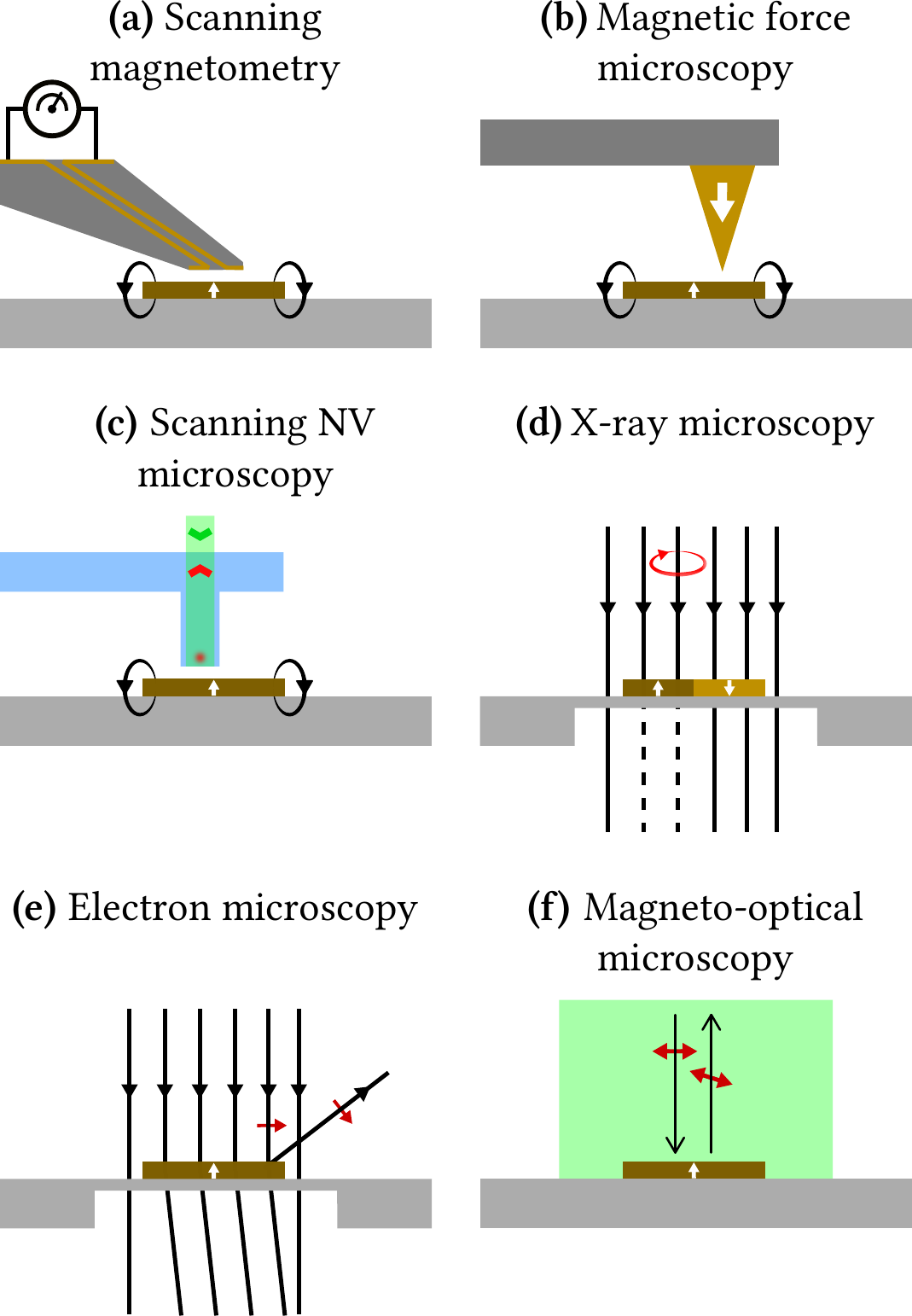}
	\caption{Schematics illustrating the principle of a few relevant magnetic microscopy techniques: (a)~Scanning magnetometry based on electrical readout (e.g.\ with a SQUID, Hall probe, or magnetic tunnel junction); (b)~Magnetic force microscopy; (c)~Scanning NV microscopy; (d)~X-ray microscopy, here full-field transmission microscopy with circularly polarized x rays; (e)~Electron microscopy, here Lorentz transmission electron microscopy (transmission) and spin-polarized low-energy electron microscopy (reflection);  (f)~Magneto-optical microscopy, here based on the Kerr effect.   
	}  
	\label{Fig_comparison}
\end{figure}

The most common method to map the stray magnetic field from a sample is to scan a magnetic field probe above it and form a magnetic image point by point. Suitable magnetic field probes include superconducting quantum interference devices (SQUIDs), Hall sensors, and magnetic tunnel junctions. These sensors rely on a magnetic-field-dependent electrical response specific to each device and therefore are capable of measuring the absolute magnetic field following calibration [Fig.~\ref{Fig_comparison}(a)]. Another type of stray field imaging technique is magnetic force microscopy (MFM), which exploits the force exerted by the stray field gradient on a magnetic tip [Fig.~\ref{Fig_comparison}(b)]. 

As is often the case in microscopy, there are tradeoffs between sensitivity and spatial resolution as they both scale with the size of the probe. SQUID microscopes offer the highest sensitivities, being able to resolve fields of order $100$\,nT at a spatial resolution of $\sim$\,100\,nm with the smallest loops\cite{vasyukovScanningSuperconductingQuantum2013,tschirhartImagingOrbitalFerromagnetism2021} (nanoSQUIDs), or down to $\sim$\,1\,nT with larger loops giving a spatial resolution of several $\mu$m.\cite{bertDirectImagingCoexistence2011,nowackImagingCurrentsHgTe2013} SQUIDs can tolerate applied fields of up to $\sim$\,1\,T.\cite{vasyukovScanningSuperconductingQuantum2013}  One drawback of SQUIDs is that they only operate at cryogenic temperatures (typically $<8$\,K). 

Hall sensors and magnetic tunnel junctions can be fabricated down to a similar size of order $100$\,nm but with far lower field sensitivity than SQUIDs, of the order of 10\,$\mu$T.\cite{oralRealTimeScanning1996,schragScanningMagnetoresistiveMicroscopy2003,kirtleyFundamentalStudiesSuperconductors2010,tangInterchangeableScanningHall2014,kogelMagneticFieldCurrent2016} Hall probes have the advantage of operating at any temperature and magnetic field, but tend to be mechanically delicate and sensitive to destruction by electrostatic discharge. Magnetic tunnel junctions are designed to operate near room temperature and in low field only (below about $10$\,mT), and are magnetically invasive to the sample. 

MFM has a spatial resolution of 10-100\,nm and is also magnetically invasive but can operate at virtually any temperature and field.\cite{kazakovaFrontiersMagneticForce2019} Even though MFM is sensitive to field gradients rather than absolute field, its sensitivity is sufficient to detect, for example, a single atomic layer of a ferromagnetic material.\cite{yuChemicallyExfoliatedVSe22019} The main drawback of MFM is difficulty in interpreting its signal, and propensity for artifacts.\cite{kazakovaFrontiersMagneticForce2019}

We also note the existence of a stray-field imaging technique that has a similar concept to widefield NV microscopy, but where the NV layer is replaced by a magneto-optically active film that is read out by standard widefield magneto-optic microscopy.\cite{indenbomVisualizationMagneticStructures2002,ueharaAdvancesMagnetoopticalImaging2010} However, the sensitivity of this technique is comparatively poor (smallest resolvable fields of order 100\,$\mu$T), the signal amplitude is difficult to calibrate, and the sensing layer is magnetically invasive.

Compared to these techniques, NV-based magnetic microscopy (whether scanning or widefield) presents a unique advantage: it is the only technique that can measure the absolute magnetic field in a calibration-free way, because it measures frequency shifts that have a simple, predictable relationship with the magnetic field [Eq.~(\ref{eq:BZ})]. This simplicity makes them ideal for experiments that require high accuracy, with some caveats discussed in Sec.~\ref{subsec:accuracy}. 

Scanning NV microscopy [Fig.~\ref{Fig_comparison}(c)] routinely offers a spatial resolution of 50\,nm (given by the NV-sample distance) and a smallest resolvable field of 10\,$\mu$T, far inferior to SQUIDs but comparable to other techniques and sufficient to detect a ferromagnetic monolayer.\cite{thielProbingMagnetism2D2019,sunMagneticDomainsDomain2021} It is also more versatile than SQUID microscopes, being able to operate in principle at any temperature (from cryogenic up to 600\,K\cite{toyliMeasurementControlSingle2012}), but has a more limited field range (typically $\le$\,0.3\,T, see Sec.~\ref{subsec:versatility}). Compared with scanning Hall probe microscopy, scanning NV microscopy has an intrinsic robustness advantage as the diamond tip is naturally suited to scanning whereas miniature Hall probes have a complex design that makes it difficult to scan a nonflat sample in close contact. As a result, scanning NV microscopy emerges as the most versatile technique available for accurate stray field mapping of a sample with high spatial resolution over a broad range of experimental conditions.       

Widefield NV microscopy has a spatial resolution limited by optics even if the sample is in close contact with the diamond. The optical diffraction limit sets the smallest achievable resolution to approximately 400\,nm but optical aberrations often spoil the resolution, as we will discuss in Sec.~\ref{subsec:spatial_res}. One advantage over scanning NV microscopy is the improved ensemble sensitivity, with fields below $100$\,nT typically resolvable.\cite{glennMicrometerscaleMagneticImaging2017,turnerMagneticFieldFingerprinting2020} This improved sensitivity is enabled in part by the parallel nature of the acquisition with a camera, rather than raster scanning a localized probe. Another advantage is the relative simplicity and robustness, with no moving parts involved when taking an image. Widefield NV microscopy is also better suited for imaging over large areas (with a field of view of up to several millimeters\cite{glennMicrometerscaleMagneticImaging2017,turnerMagneticFieldFingerprinting2020}), whereas scanning NV microscopy is typically limited to a few tens of microns only. 

Widefield NV microscopy is also the only stray-field technique that enables vector magnetometry in routine use. All other techniques mentioned above, including scanning NV microscopy, are by default sensitive to a single component of the magnetic field and extension to vector sensing remains a challenge.\cite{anahoryThreeJunctionSQUIDonTipTunable2014} Although full vector information in a 2D magnetic field image in general does not provide critical additional information compared to measuring a single projection,\cite{casolaProbingCondensedMatter2018,limaObtainingVectorMagnetic2009} it has been shown that vector information can improve the accuracy of the reconstruction of current distributions.\cite{broadwayImagingDomainReversal2020} The large field of view also contributes to facilitating accurate reconstructions of current or magnetization distributions by reducing truncation artifacts that plague high-resolution techniques like scanning NV microscopy.\cite{broadwayImagingDomainReversal2020,thielProbingMagnetism2D2019}

Stray field mapping techniques based on SQUIDs, Hall probes or magnetic tunnel junctions can be scaled to allow imaging over large areas up to centimeters with a spatial resolution down to a few microns,\cite{gregoryScanningHallProbe2002,kirtleyFundamentalStudiesSuperconductors2010} thus offering similar capabilities to widefield NV microscopy. In comparison, NV microscopy then has the advantage of a better sensitivity (over Hall probes and magnetic tunnel junctions) only inferior to SQUIDs.\cite{fuHighSensitivityMomentMagnetometry2020} Consequently, widefield NV microscopy appears to be most useful for magnetic surveys of weak magnetic features (of the order of 1\,$\mu$T or less) when room temperature operation (or variable down to cryogenic) is required. Examples of recently demonstrated applications that meet these criteria are: the detection of weak magnetic moments in geological samples,\cite{fuHighSensitivityMomentMagnetometry2020} the monitoring of integrated-circuit activity,\cite{turnerMagneticFieldFingerprinting2020} and the study of ultrathin ferromagnetic materials.\cite{broadwayImagingDomainReversal2020} 

Nevertheless, widefield NV microscopy currently faces some limitations that hamper broader applicability, which will be discussed in detail in Sec.~\ref{sec:limitations}. One intrinsic shortcoming is limited spatial resolution which may prevent in-depth investigations of some samples and phenomena, such as observing the domain structure of thin magnetic films\cite{broadwayImagingDomainReversal2020} or mapping fine details of current distributions.\cite{kuImagingViscousFlow2020} As such, we see widefield NV microscopy as a complement of higher-resolution techniques such as scanning NV microscopy, which could be used to ``zoom in'' on a pre-identified region of interest, albeit with an inferior sensitivity. 

\subsubsection{Interaction techniques}

Many different methods exist to probe micron-sized ferromagnetic samples by interacting test particles (e.g.\ photons or electrons) with the sample in a transmission or reflection configuration, a review of which can be found in Ref.~\onlinecite{fischerXRayImagingMagnetic2015}. Each technology's applicability often depends on the type of material and sample geometry considered.

For magnetic thin films, several techniques are available that offer a spatial resolution of 10\,nm or less. Most relevant to our discussion are transmission X-ray microscopy\cite{fischerXRayImagingMagnetic2015} [Fig.~\ref{Fig_comparison}(d)] and various types of electron microscopy including Lorentz transmission electron microscopy\cite{yuRealspaceObservationTwodimensional2010} and spin-polarized low-energy electron microscopy\cite{chenTailoringChiralityMagnetic2013} [Fig.~\ref{Fig_comparison}(e)]. A common limitation to these high lateral spatial resolution techniques, however, is their limited sensitivity that is insufficient to image, for example, ferromagnetic films in the ultrathin limit (a few atomic layers). As such, widefield NV microscopy, with its high sensitivity, is particularly interesting for investigations of atomically thin magnetic structures.\cite{broadwayImagingDomainReversal2020} 

For such ultrathin samples, the technique most comparable to widefield NV microscopy is magneto-optical imaging [Fig.~\ref{Fig_comparison}(f)], which exploits magnetic birefringence or dichroism and manifests in a rotation of the polarization of linearly polarized incident light (Kerr effect) or in a change in its ellipticity (in magnetic circular dichroism measurements).\cite{makProbingControllingMagnetic2019} These techniques are routinely employed to study ultrathin sputtered films\cite{mihaimironCurrentdrivenSpinTorque2010} and recently have become the methods of choice to investigate van der Waals 2D magnets.\cite{makProbingControllingMagnetic2019} Like widefield NV microscopy, the spatial resolution is limited by optical diffraction and the temperature range extends from cryogenic to room temperature. A convenient feature of magneto-optical imaging is that the range and direction of external magnetic field is not limited which makes it possible to record spatially resolved hysteresis curves even for high coercivity materials and along any direction. A limitation of this technique is that it is only applicable to materials with accessible optical transitions and a sufficiently large spin-orbit coupling.\cite{makProbingControllingMagnetic2019}

In contrast, widefield NV microscopy is typically limited to fields of less than 0.3\,T applied along specific directions (see Sec.~\ref{subsec:versatility}), but has a few advantages. First, it is universally applicable to any material that produces a stray field, including ferromagnets with weak spin-orbit coupling, paramagnets and diamagnets. Moreover, unlike magneto-optical imaging or the higher resolution techniques mentioned above, NV microscopy is inherently quantitative without any calibration. This enables the absolute magnetization (spontaneous or induced) of a sample to be spatially mapped, although some modeling and assumptions are necessary.\cite{broadwayImagingDomainReversal2020} This calibration-free, quantitative nature is also advantageous when studying the effects of, for example, temperature, surface or interface effects, strain, doping, or current injection.\cite{gongTwodimensionalMagneticCrystals2019} Indeed, in these instances the magneto-optical response of the sample may change independently of changes in the magnetization of the magnetic film, whereas a change in stray field necessarily indicates a change in the magnetic properties of the sample. 

For more complex three-dimensional magnetic structures such as nanoparticles, interaction techniques working in transmission, e.g.\ electron holography,\cite{dunin-borkowskiOffaxisElectronHolography2004,shindoElectronHolographyMagnetic2008} are commonly employed. However, these techniques are not generally applicable to thicker objects (typically larger than 100\,nm). In particular, magnetic materials embedded in a thick solid matrix (e.g.\ a slice of a rock or animal) are best probed using stray-field techniques. In this context, widefield NV microscopy is a leading technique because it provides the best sensitivity among stray-field techniques that operate at ambient conditions.\cite{mccoeyQuantumMagneticImaging2020,fuHighSensitivityMomentMagnetometry2020} One such application is the analysis of magnetic inclusions in rocks, allowing both the direction and magnitude of magnetic moments to be accurately determined.\cite{fuHighSensitivityMomentMagnetometry2020}

Finally, there is a limited number of techniques able to probe current distributions without measuring the \O rsted magnetic field. These include scanning gate microscopy\cite{bhandariImagingCyclotronOrbits2016} and Hall field imaging using a single-electron transistor.\cite{sulpizioVisualizingPoiseuilleFlow2019} Both of these techniques have limited applicability (for instance, buried structures are out of reach). Thus, stray magnetic field microscopy remains the most versatile method for spatially mapping current distributions, with the advantages of widefield NV microscopy outlined in the previous section.

\subsection{Imaging of time-varying magnetic signals}

As discussed in Sec.~\ref{subsec:ac}, widefield NV microscopy offers the possibility to image AC/fluctuating magnetic signals at kHz to GHz frequencies,\cite{levinePrinciplesTechniquesQuantum2019} as well as to stroboscopically image transient phenomena with a 10 nanosecond temporal resolution. Importantly, these measurements are often quantitative, i.e.\ the amplitude of the transient or AC magnetic field, or the spectral noise density in the case of fluctuating signals, can be determined.

Scanning magnetometry techniques (based on SQUIDs, Hall probes, magnetic tunnel junctions) can readily measure kHz signals, but accessing larger frequencies remains a challenge. On the other hand, several of the interaction techniques described in the previous section are also capable of accessing GHz signals, albeit in a qualitative manner (in the sense that the amplitude of the signal cannot be readily related to material properties). For instance, x-ray transmission microscopy and magneto-optical microscopy can image (stroboscopically) changes in the magnetic texture of a thin film with a 100 picosecond temporal resolution.\cite{vanwaeyenbergeMagneticVortexCore2006,xuMagnetizationDynamicsVortexImprinted2012} These techniques, as well as Brillouin light scattering microscopy, are also able to image the modes and propagation of spin waves in a magnetic thin film.\cite{sebastianMicrofocusedBrillouinLight2015,ursAdvancedMagnetoopticalMicroscopy2016,slukaEmissionPropagation1D2019}

Nevertheless, NV microscopy offers unique opportunities for quantitative studies of time-varying magnetic phenomena, such as spin waves.\cite{bertelliMagneticResonanceImaging2020} Moreover, the techniques mentioned above have limited sensitivity, such that minute fluctuating magnetic signals from nonferromagnetic samples, e.g.\ electron and nuclear spin ensembles, are typically out of reach. In this regime, we note the existence of another technique that enables spatially resolved EPR and NMR, magnetic resonance force microscopy.\cite{poggioForcedetectedNuclearMagnetic2010} However, this technique is experimentally challenging, requiring low (sub-Kelvin) temperature. Thus, the possibility of performing in-situ EPR/NMR characterisation of a sample within a widefield NV microscope opens unique prospects, motivating further work to improve the spectral resolution (especially for NMR) and hence chemical sensitivity.\cite{ziemQuantitativeNanoscaleMRI2019}

\subsection{Summary and outlook}

Based on the above discussions, we conclude that two of the most important strengths of widefield NV microscopy are its high sensitivity and its quantitative nature. Therefore, the most promising applications will likely be those that directly benefit from these competitive advantages. Examples mentioned previously are: (i)~the quantitative analysis of small amounts of magnetic materials embedded in a larger nonmagnetic host;\cite{lesageOpticalMagneticImaging2013,mccoeyQuantumMagneticImaging2020,fuHighSensitivityMomentMagnetometry2020} and (ii)~the precise monitoring of the magnetization of 2D magnetic materials under various stimuli and environmental changes.\cite{broadwayImagingDomainReversal2020} Attractive additional features of widefield NV microscopy are its relative versatility (temperature range, applicability to any material) and the possibility to exploit other sensing modalities \textit{in situ}, e.g.\ chemical analysis via NMR spectroscopy. 

However, we also highlighted some limitations such as spatial resolution and applicable magnetic field, which means widefield NV microscopy appears at this stage to be most useful as a complement to other techniques rather than as a primary technique. For instance, in the case of thin-film studies, magneto-optical microscopy is preferred (given its simplicity and higher flexibility in terms of applied field) when an absolute measurement of the magnetization is not required, whereas scanning NV microscopy is the method of choice when a higher spatial resolution is necessary. These considerations motivate further development of widefield NV technology to make it more versatile and easy to use and raise its prospects as a primary microscopy technique.

\section{\label{sec:limitations}Current limitations and future improvements} 

Here we describe in more detail some of the limitations and challenges faced by widefield NV microscopy, and identify directions for future work that could alleviate those limitations.

\subsection{\label{subsec:spatial_res}Spatial resolution} 

Spatial resolution in widefield NV microscopy is limited by three contributing factors: the optical components, NV-sample standoff, and thickness of the NV layer.\cite{levinePrinciplesTechniquesQuantum2019} The optical limit is imposed by the optics of the system and bounded from below by the diffraction limit ($\approx$\,$400$\,nm with a high numerical aperture objective). Reaching the diffraction limit is generally prevented by aberrations from the diamond and lenses. Standoff between the sample and NV layer introduces a convolution of the stray-field at the NV layer with a Lorentzian, broadening features and inhibiting the ability to resolve small, tightly packed magnetic objects such as magnetic domains. NV layer thickness further contributes to this by introducing variation in the distance from the nearest NVs and those furthest from the sample.

\begin{figure*}[tb]
	\includegraphics[width=0.99\textwidth]{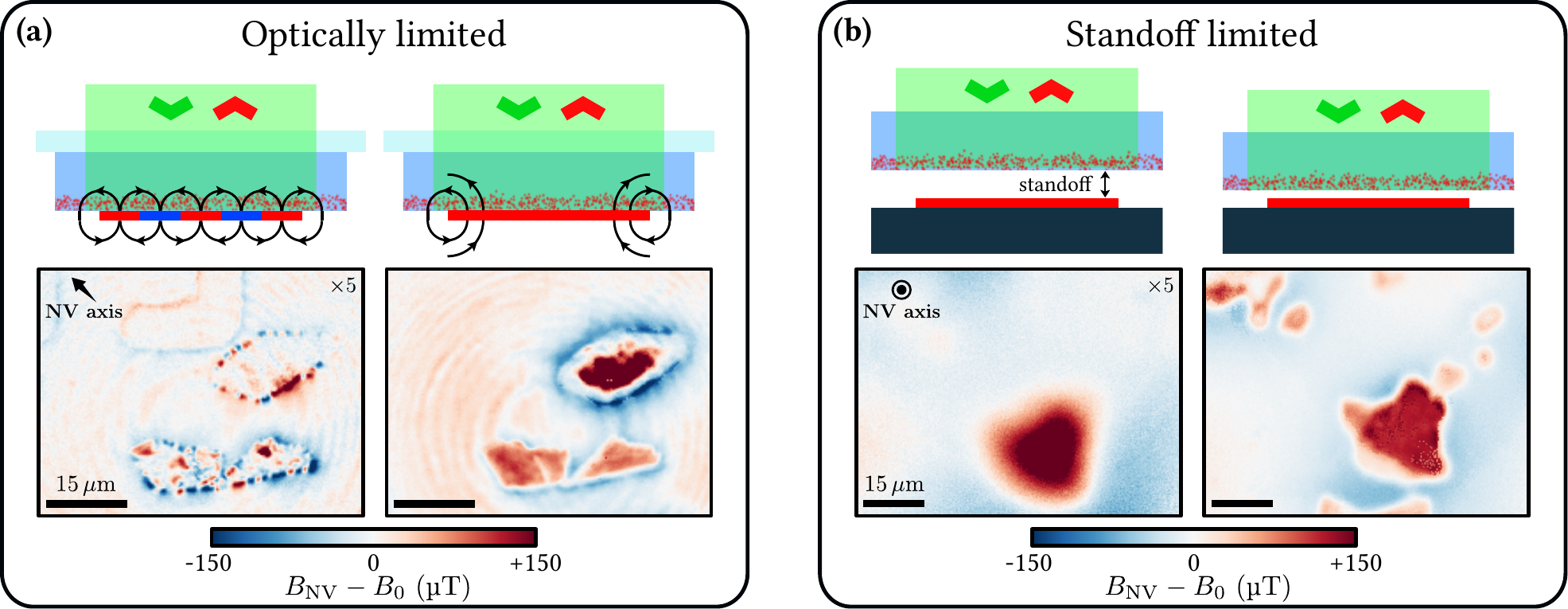}
	\caption{Impacts of optical and standoff limitations on the spatial resolution, illustrated by images of Fe$_3$GeTe$_2$ magnetic flakes ($<$\,100\,nm thickness). (a)~Magnetic field maps of the same flakes after zero-field cooling (left) and after magnetization (right), showing how the interior of the flake lacks defined structure before magnetization as most domains are unresolved if they are smaller than the optical resolution limit. Images are taken using a sample-on-diamond device, with a \{100\}-oriented diamond and combined standoff and NV layer thickness of $\approx$\,300\,nm. A bias magnetic field of $B_0=6$\,mT was applied to take the images. (b)~Magnetic field maps of magnetized flakes imaged with different standoff distances. At larger standoff distances there is a distinct loss of well defined features. Images are taken with a diamond-on-sample device using a \{111\}-oriented diamond. $B_{\rm NV}$ is equivalent to $B_Z$ in Eq.~(\ref{eq:BZ}).}
	\label{fig:spatial_res}
\end{figure*}

We demonstrate the impacts of spatial resolution limitations on widefield images in the two cases of diamond-on-sample and sample-on-diamond interfacing. The test sample we use is thin exfoliated flakes of Fe$_3$GeTe$_2$, a magnetic van der Waals material; the experimental methods are the same as in Ref.~\onlinecite{broadwayImagingDomainReversal2020}. Figure~\ref{fig:spatial_res}(a) (left) shows an unmagnetized, zero-field cooled flake on a sample-on-diamond device. Under these conditions the interior of the flake is expected to exhibit some disorganized domain structure, however, these features are not resolved as they are below the optical resolution limit (here about 1\,$\mu$m). Instead we see the averaged signal which is no net magnetic field. Once magnetized with a large bias field ($B_z = 1$\,T) into a single domain, the interior structure of the flake is revealed and the amplitude of the magnetization can be inferred [Fig.~\ref{fig:spatial_res}(a), right]. For the sample-on-diamond device, the standoff distance is fixed by design. Conversely, the diamond-on-sample device requires the diamond to be moved into position on a potentially rough surface which may incur an undesirable standoff distance. Figure~\ref{fig:spatial_res}(b) shows the difference between images taken with two different standoffs. When the diamond is brought in close contact with the sample (through trial and error), the flakes are better resolved and a defined structure is visible. 

Optimizing spatial resolution imposes restrictions on the standoff and thickness of the NV layer. For example, the diffraction limit for a widefield system is typically $\approx$\,400\,nm. To be imaging at this limit, the combined distance of the standoff and NV layer thickness from the sample must fall below this value. This is due to the full width at half maximum of the convolved Lorentzian scaling proportionally with this distance. Generally, sensitivity is improved by increasing the thickness of the NV layer and minimizing the standoff distance; implying a trade-off between spatial resolution and sensitivity. In the above case of trying to image at the diffraction limit, it is best to tailor the standoff and NV layer thickness to maximize sensitivity whilst still being below the diffraction limit. For example, the images in Fig.~\ref{fig:spatial_res}(a) were obtained using a diamond with a combined standoff and NV layer thickness of $\approx$\,300\,nm. 

To reach the diffraction limit, optical components should be chosen to minimize aberrations in the PL signal as it passes through the diamond and subsequent optics along the path to the camera. For instance, the optical resolution of 1\,$\mu$m in Fig.~\ref{fig:spatial_res}(a) is likely a result of aberrations caused by imaging through the 100\,$\mu$m diamond. Given how critical the spatial resolution is in many applications [e.g.\ to resolve the domain structure in Fig.~\ref{fig:spatial_res}(a)],  improving the optical design to minimize optical aberrations in widefield NV microscopy will be an important direction of future research.

\subsection{\label{subsec:accuracy}Accuracy} 

\begin{figure*}[tb]
	\includegraphics[width=0.99\textwidth]{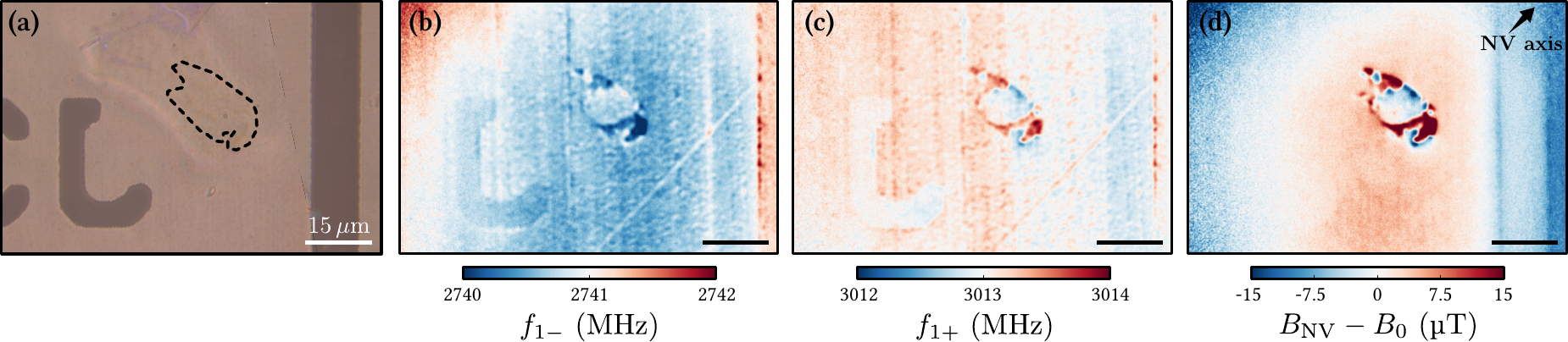}
	\caption{Systematic errors in widefield NV microscopy. (a)~Optical micrograph of region imaged in~(b-d), showing metallic registration marks (dark: diamond only; light: metal coated). The magnetic sample, an ultrathin flake of van der Waals ferromagnet VI$_3$ (3 atomic layers thick, or $\approx$\,2\,nm) is outlined. (b),(c)~Maps of the resonance frequencies $f_{1-}$ (b) and $f_{1+}$ (c) recorded for an NV layer ($\approx$\,100\,nm thick)  under an aligned bias field of amplitude $B_0=5$\,mT. These maps are dominated by strain features of various length scales. (d)~Magnetic field map obtained by subtraction of (b) from (c) via Eq.\,(\ref{eq:BZ}) (we rename $B_Z$ as $B_{\rm NV}$ to avoid confusion with the lab frame $z$), revealing the stray field from the VI$_3$ flake. Figure adapted with permission from Ref.~\onlinecite{broadwayImagingDomainReversal2020}. Copyright 2020, Wiley-VCH GmbH.}
	\label{fig:strain}
\end{figure*}

Being a quantitative technique, widefield NV microscopy is often used for estimating source quantities, for example macroscopic magnetic moments,\cite{fuHighSensitivityMomentMagnetometry2020} and current densities or magnetizations in thin films\cite{tetienneQuantumImagingCurrent2017,broadwayImagingDomainReversal2020} may be calculated algebraically from magnetic field images. However, these reconstruction algorithms are achieved via upward continuation Fourier space methods that are highly unstable to noise or inaccuracy in the magnetic field map input.\cite{meltzerDirectReconstructionTwoDimensional2017,feldmannResolutionTwodimensionalCurrents2004,clementReconstructionCurrentDensities2019} For example, we have cataloged a range of artifacts these instabilities introduce for thin film sources.\cite{broadwayImagingDomainReversal2020} In this regime, further work benchmarking widefield NV microscopy against a reference technique or sample (similar to previous work done for magnetic moments\cite{fuHighSensitivityMomentMagnetometry2020}) is desirable. 

In addition to reconstruction errors, there are multiple common sources of inaccuracy in the static magnetic imaging process itself. Here we will group the various effects into 3 classes: fundamental limitations, complex ODMR spectra and delocalized signal. The first limitation has already been mentioned: at low magnetic field the perturbations of strain, electric and magnetic fields on the NV are not completely separable in general, even if the resonances are perfectly resolved.\cite{broadwaySpatialMappingBand2018,broadwayMicroscopicImagingStress2019} As a result, internal electric field and strain variations within the NV layer\cite{broadwaySpatialMappingBand2018,kehayiasImagingCrystalStress2019} can cause artifacts in a magnetic field map measured at a low bias magnetic field. Sensitivity to strain and electric fields can be reduced by applying a sufficiently large bias field ($B_0\gtrsim5$\,mT) carefully aligned to an NV axis. In this case, the resonance frequencies for the aligned NVs are 
\begin{align} \label{eq:f_strain}
f_{i\pm} &= D+a_1\Sigma_Z^{\rm axial}+2a_2\Sigma_{Z_i}^{{\rm shear}}\pm\gamma_{\rm NV}B_{Z_i}~,
\end{align}
so that the magnetic field projection can be retrieved without error by taking the difference of the frequencies, Eq.~(\ref{eq:BZ}). However, this alignment requirement hinders vector magnetometry. Moreover, under a large aligned bias field one is often able to measure only one resonance due to technical constraints; the other resonance lies in an inaccessible frequency range. In this case, strain is a direct source of error as evident from Eq.~(\ref{eq:f_strain}). 

The importance of strain is illustrated in Fig.~\ref{fig:strain}. Here the diamond is partly covered with metallic registration marks [Fig.~\ref{fig:strain}(a)] and the NV layer is about 100\,nm thick. The $f_{1\pm}$ maps, recorded under an aligned bias field $B_0=5$\,mT, are shown in Figs.~\ref{fig:strain}(b) and \ref{fig:strain}(c). The frequency variations visible on a length scale of a few microns are attributed to strain native to the diamond (possibly related to surface roughness), whereas the larger features correlating with Fig.~\ref{fig:strain}(a) correspond to strain induced by the deposited metallic structures. These various strain features are efficiently removed in the frequency difference map [Fig.~\ref{fig:strain}(d)], revealing the magnetic field from the sample (here a thin van der Waals ferromagnetic flake). This observation highlights the need for proper normalization, as strain can otherwise obscure magnetic features or introduce errors that prevent accurate reconstruction. It also motivates further efforts to minimize internal strain and devise normalization methods adapted to measurements where strain cannot be removed by simple subtraction (e.g.\ in near-zero bias field).

Another common class of inaccuracies is encountered as complex ODMR spectra. When the resonant lineshape strays from a simple Lorentzian or Gaussian, in particular when asymmetric, it can be difficult to systematically determine the resonance frequency. Nonuniform microwave driving power across the frequency sweep -- e.g.\ due to the presence of metallic objects as seen in Fig.~\ref{fig:strain}(a) -- can easily result in complex lineshapes. Charge state fluctuations or laser repumping effects during acquisition can be largely resolved by taking a reference measurement without a microwave pulse, but such normalization is not perfect and doubles the measurement time. Importantly, systematic errors due to microwave and laser effects will likely vary across the field of view, leading to imaging artifacts such as low spatial frequency modulations. These effects likely explain the residual errors seen in Fig.~\ref{fig:strain}(d) in the form of low spatial frequency background modulations as well as finer features. Similar artifacts are seen in Fig.~\ref{fig:spatial_res}(a) in the form of rings which correlate with the laser intensity profile.

Another example of an asymmetric lineshape emerges at low sample-NV standoffs.\cite{tetienneProximityInducedArtefactsMagnetic2018} This artifact is caused by the wide range of magnetic field strengths experienced by the NVs within each readout volume (pixel) combined with a loss of information by the Zeeman shifts beyond the measurement range. The resultant  magnetic field maps can deviate substantially in shape, amplitude and even sign.\cite{tetienneProximityInducedArtefactsMagnetic2018} Increasing the NV-sample standoff to $\ge$\,100\,nm -- e.g.\ by adding a spacing layer -- can be employed to avoid this issue. Additionally, spatial averaging over the local electric field environment has been shown to produce an asymmetric lineshape at near-zero bias field.\cite{mittigaImagingLocalCharge2018,lillieLaserModulationSuperconductivity2020} 

If the physical nature of an anomalous lineshape is known, a better fit model might be employed. However, each effect is not mutually exclusive with the others and/or could only appear within particular regions of an image and go undetected. Additional free parameters may be used to account for the anomalous effect in each pixel, but this is complicated by overfitting and the difficulty in keeping model parameters smooth (i.e.\ physical) across the field of view. The ubiquity of this class of inaccuracy motivates future efforts at characterization and mitigation, in particular via machine learning approaches to model selection.

Finally, the widefield approach assumes all PL imaged comes directly from the NV layer. However, reflections at the diamond surfaces lead to a delocalized contribution to the ODMR spectrum. Initial investigations by Fu et al.\cite{fuHighSensitivityMomentMagnetometry2020} found that this background signal was up to 50\% of the spectra at each pixel. Any global contribution to the ODMR spectrum must be subtracted or risk nonquantitive field measurements, yet it cannot be determined independent of the sample under test. Further work is urgently required to characterize and mitigate this issue in a systematic manner.

\subsection{\label{subsec:field_sens}Sensitivity and materials considerations}

The high magnetic sensitivity of widefield NV microscopy is one of the strengths of this technology, with fields as small as 100\,nT sometimes resolvable.\cite{glennMicrometerscaleMagneticImaging2017} However, reaching this level of sensitivity routinely is not straightforward.

Two main factors limit the sensitivity in widefield NV microscopy. The first one is related to accuracy. As discussed in the previous section, systematic errors that vary across the field of view may obscure localized contributions from the sample. These background spatial variations can often be removed by normalization (e.g.\ by frequency subtraction or bias field modulation), but residual errors typically remain. For instance, in Fig.~\ref{fig:strain}(d) background variations of order 10\,$\mu$T are visible, which may limit the ability to detect a sample's signal at this level if the length scales are similar. 

The second factor affecting sensitivity is noise, as for any sensor. For a shot-noise limited pulsed ODMR measurement, this can be seen through the equation for DC magnetic sensitivity:\cite{rondinMagnetometryNitrogenvacancyDefects2014,barrySensitivityOptimizationNVdiamond2020} 
\begin{align}
\eta_{\rm DC} = \frac{1}{\gamma_{\rm NV} \mathcal{C}T_2^*\sqrt{\alpha \mathcal{R}}},
\label{eq:sensitivity}
\end{align}
where $T_2^*$ is the free induction decay time of the NV spins, $\mathcal{C}$ is the absolute spin contrast, $\mathcal{R}$ is the NV photon count rate during readout, and $\alpha$ is the readout duty cycle. This shot-noise sensitivity is dependent on the measurement conditions and the setup used as $\alpha$ is chosen, in practice, to achieve a balance between maximizing $\mathcal{C}$ and ensemble repumping, and $\mathcal{R}$ is dependent on the photon collection efficiency. In conditions typical for widefield NV microscopy, sensitivities of order $\mu$T/$\sqrt{\rm Hz}$ over a diffraction-limited area are common with current technology,\cite{healeyComparisonDifferentMethods2020} in principle allowing the detection of signals down to 100\,nT scale in minutes. However, quantitative, properly normalized images require measurement times on the order of hours to reach this sensitivity level. Additionally, in practice the noise floor is often limited by technical factors and experimental drifts, in some cases preventing the shot-noise limit from being achieved. For instance, the magnetic field map in Fig.~\ref{fig:strain}(d) has a background noise (pixel-to-pixel noise) of 1.5\,$\mu$T, obtained after over 10 hours of integration. We note that the advent of lock-in cameras may facilitate reaching the shot-noise limit by removing noise up to $\sim$\,kHz frequencies,\cite{wojciechowskiContributedReviewCameralimits2018,hartDiamondMagneticMicroscopy2021,webbHighSpeedMicrocircuit2021,parasharLockinDetectionBased2021} whereas standard cameras are limited to modulation at the camera frame rate of typically 30\,Hz. Regardless, the magnetic sensitivity of NV ensembles still requires optimization in order to routinely meet the 100\,nT benchmark and reduce measurement times.  

Straightforwardly, it can be seen from Eq.~(\ref{eq:sensitivity}) that it is beneficial to maximize both $T_2^*$ and $\mathcal{R}$, but these quantities generally compete against each other as they scale inversely with the N density of the diamond.\cite{bauchDecoherenceEnsemblesNitrogenvacancy2020} For widefield imaging, Eq.~(\ref{eq:sensitivity}) can usually be simplified by taking $\alpha \approx 1$ as low laser power densities often necessitate the use of long readout pulses. It can be seen, then, that the sensitivity varies as $T_2^*\sqrt{\mathcal{R}}$, which implies that NV layers with a low N density (long $T_2^*$, small $\mathcal{R}$) are generally more sensitive than with larger N density (short $T_2^*$, large $\mathcal{R}$), assuming the N-to-NV yield is constant.\cite{healeyComparisonDifferentMethods2020,barrySensitivityOptimizationNVdiamond2020} Note that this is true provided the nitrogen spin bath is the dominant source of NV decoherence. In the lower density regime, other independent effects can have a particularly harmful effect on sensitivity as they are not offset by high $\mathcal{R}$. These include the presence of crystal strain broadening ODMR linewidths\cite{hartDiamondMagneticMicroscopy2021} and, where present, the $^{13}$C bath.\cite{bauchDecoherenceEnsemblesNitrogenvacancy2020} As discussed in Sec.~\ref{subsec:sensor_design}, the CVD growth method is the most suited to forming NV layers of arbitrary thickness and N density. Commercial availability of NV-diamonds made with this method is therefore highly desirable.

Regardless of the creation method, a source of improvement may come from increasing NV yield. Current implantation and annealing procedures typically result in less than 5\% N-to-NV conversion,\cite{healeyComparisonDifferentMethods2020} compounded by the fact that only 25\% of NVs can be addressed at once. Strategies for improving conversion efficiency to near unity, based on lattice charging via the introduction of shallow donors, have recently been demonstrated on low-density NV ensembles;\cite{luhmannCoulombdrivenSingleDefect2019} however, their success in the high-nitrogen regime is yet to be demonstrated. In-situ annealing during irradiation/implantation is another possible strategy for improving yield, yet to be explored for thin NV layers.\cite{kucskoCriticalThermalizationDisordered2018} 

In practice, maximizing $\mathcal{R}$ over $T_2^*$ may be favored when the photon collection efficiency on a particular setup is poor, or in a measurement with a low readout duty cycle. In both cases, dark counts from the camera used may dominate over shot noise if the NV PL is not high enough. In this regime, the HPHT method with its high NV density (despite shorter $T_2^*$) is advantageous. However, at very high N densities ($>$\,100\,ppm) electron tunneling between NV centers and nearby substitutional nitrogen reduces NV spin polarization\cite{mansonNVPairCentre2018} and so sensitivity in this regime will be reduced. This tunneling process may place an upper limit on the brightness of NV ensembles. 

Finally, experimental convenience often requires an NV ensemble to have consistent properties (brightness, coherence, etc) across the measurement field of view. While naturally achieved in CVD growth (up to  macroscopic growth defects/features), HPHT growth is prone to the incorporation of different growth ``sectors'' with differing nitrogen densities. This inhomogeneity can often be worked around by pre-screening the diamond, however it is not easy to avoid completely when using commercial substrates, and thus adds an unwanted complication to the method that best lends itself to high-throughput measurements due to its low cost. Substrate variability and inhomogeneity could be mitigated in the future if the production and sale of HPHT substrates designed for this purpose become commonplace. 

Although this discussion has been confined to DC magnetic sensitivity, these concerns generalize to the measurement of other target fields. For example, in the case of AC magnetometry, it is usual to modify Eq.~(\ref{eq:sensitivity}) to represent a figure of merit for AC sensitivity by replacing $T_2^*$ by the coherence time $T_2$ specific to the sequence used.\cite{rondinMagnetometryNitrogenvacancyDefects2014,barrySensitivityOptimizationNVdiamond2020,taylorHighsensitivityDiamondMagnetometer2008}

\subsection{\label{subsec:versatility}Operating conditions}

\subsubsection{Temperature}

One of the strengths of NV sensing is the wide range of environments in which it is sensitive without re-calibration. In particular, NVs are theoretically addressable for all temperatures up to 600\,K where nonradiative processes diminish spin selectivity.\cite{toyliMeasurementControlSingle2012} However, precise control of the sample's temperature is not straightforward due to the invasiveness of the ODMR technique both from laser and microwave.

Efficient initialization of the NV requires a laser power density as near to the optical cycling saturation limit (1\,MW$/{\rm cm}^2$) as possible. If absorbed by the sample, such a power density can lead to substantial heating, with 30\,K heating above ambient reported.\cite{glennMicrometerscaleMagneticImaging2017} Microwave driving can also provide heat to the sample, indicating a limitation on high Rabi frequency experiments.

These effects must be carefully characterized and monitored especially when imaging samples in low-temperature phases, which is usually the motivation for imaging in a cryostat. Possible mitigation techniques include side illumination with angle-polished diamonds to achieve total internal reflection from the diamond surface closest to the sample, metalized diamond surfaces and improved heat-sinking. Despite a metalized diamond surface, we have found that a modest illumination power of 2\,mW (40\,${\rm W}/{\rm cm}^2$) can heat the sample locally by 5\,K above a 4\,K base temperature.\cite{lillieLaserModulationSuperconductivity2020} This observation also suggests that reaching sub-Kelvin temperatures will be challenging. Particularly sensitive samples might be imaged in a raster fashion by scanning a confocal spot to minimize total laser power, at the penalty of increased imaging time.\cite{bertelliMagneticResonanceImaging2020}

\subsubsection{Bias magnetic field}

NV magnetometry measurements are often performed  under a small bias magnetic field for convenience (e.g.\ 5\,mT as in Fig.~\ref{fig:strain}) but zero bias field measurements are also possible for very sensitive samples.\cite{glennMicrometerscaleMagneticImaging2017} On the other hand, analysis of magnetic samples often benefits from probing with large magnetic fields applied in arbitrary directions (e.g.\ in plane and out of the plane for a thin film). Ferromagnets with a large coercivity (hard magnets) can require up to several teslas to map out their full hysteresis curve. While it is possible to return to low bias fields to measure after polarizing the sample with a large field pulse,\cite{broadwayImagingDomainReversal2020} it is preferable in many cases (e.g.\ paramagnetic samples) to be able to image directly in the conditions of the pulse. 

However, the photophysics of the NV center imposes some restrictions on the magnetic field that can be applied during a measurement. Any off-axis magnetic field beyond about 5\,mT magnitude induces spin mixing that drastically reduces ODMR contrast.\cite{tetienneMagneticfielddependentPhotodynamicsSingle2012} Consequently, measurements in a large field require that the field be aligned with one of the four available NV axes, which is tied to the crystal orientation. For instance, with a standard \{100\}-oriented diamond, the field will form an angle of $54.7^\circ$ with the out-of-plane direction ($z$ axis). Moreover, only a single NV orientation can be measured in a given field direction, precluding vector magnetometry. 

This alignment constraint requires special hardware to allow a large magnetic field to be precisely aligned with an NV axis, such as a superconducting vector magnet. Additionally, as bias fields (and thus resonance frequency) increase the demands on microwave delivery hardware increase as well. Moderate fields up to about \ 0.3\,T can be accommodated by relatively standard microwave electronics as it corresponds to a lower resonance frequency $f_{i-}$ of about 6\,GHz. However, already at this field the higher resonance becomes difficult to access ($f_{i+}\sim11$\,GHz),\cite{fescenkoDiamondMagneticMicroscopy2019} noting that single-frequency images are prone to artifacts from common mode shifts (see Fig.~\ref{fig:strain}).  

Performing widefield NV experiments in fields larger than 0.3 T will require significant technical developments to enable delivery of high-frequency microwaves to the sample,\cite{aslamSingleSpinOptically2015} or fast field pulsing methods to return the NVs to a low field during the optical readout.\cite{bodenstedtNanoscaleSpinManipulation2018} Additionally, to enable measurements under large out-of-plane or in-plane bias fields (as desirable for thin-film studies), the availability of less common \{111\}- and \{110\}-oriented diamonds will need to be addressed. Ideally, a versatile widefield NV microscope would offer the possibility to seamlessly switch between different diamond crystal orientations so that the bias field direction most relevant to the sample under study can be applied.

\subsection{Ease of use}

To date, studies utilizing widefield NV microscopy have largely treated the NV microscope as an experiment in and of itself, in addition to the particular sample under examination. In particular, each ``experiment'' typically involves a significant amount of time dedicated to interfacing the diamond sensor with the sample (described in Sec.~\ref{subsec:interfacing}). Even the seemingly simpler diamond-on-sample approach still requires considerable know-how and patience to clean the diamond surface, position it onto the sample, make manual adjustments until the desired standoff distance is achieved, and lock the final position in (e.g.\ with glue) to enable stable measurements for an extended period of time. The last step is especially critical for cryogenic experiments, as a loosely mounted diamond tends to move upon pumping and cooling. This requirement of specialist expertise is a major barrier to wider adoption of widefield NV microscopy.  

\begin{figure}
	\includegraphics[width=0.45\textwidth]{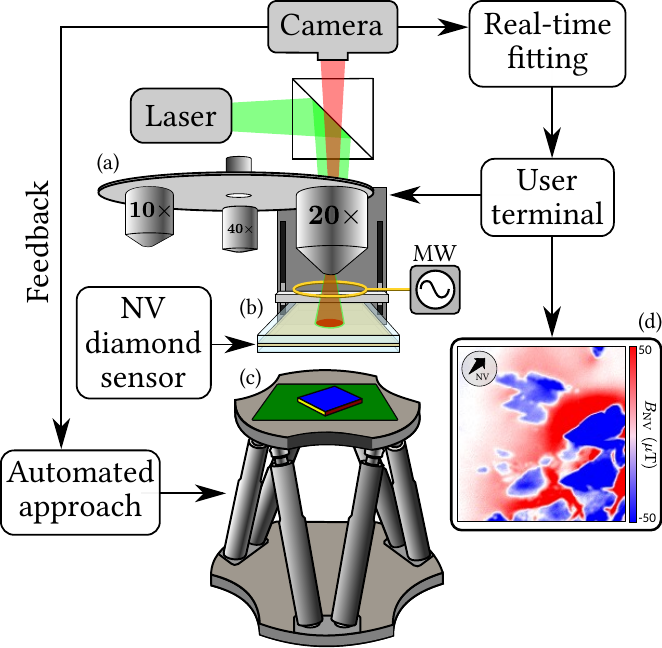}
	\caption{Illustration of a widefield NV microscope implementing features to support streamlined user operation. Objectives of various magnification are mounted to a turret (a) for rapid interchange. The diamond sensor (b) is mounted separately to the sample (c), so that different samples can be imaged without affecting the sensor configuration. The sample (c) sits on a stage with orientation control (e.g.\ a hexapod). Together with feedback from the camera, this enables ``automated approach'': automatically bringing the sample and sensor surfaces into alignment and contact. Once sensor-sample alignment is completed, the user can view the magnetic image in real-time (d) e.g.\ using GPU accelerated data processing. Components not to scale.}  
	\label{fig:easeofuse}
\end{figure}

In this section, we discuss a few directions for future work to overcome current challenges pertaining to usability in operation and flexibility of application, including this interfacing issue. The proposed technological developments are illustrated in Fig.~\ref{fig:easeofuse}.

\subsubsection{Simpler interfacing}

As discussed in section~\ref{sec:implementation}, the two approaches to NV-sample interfacing are sample-on-diamond involving sample fabrication on the diamond sensor, and diamond-on-sample where the sample and diamond are brought together for measurement only. A general purpose widefield NV microscope necessitates the second approach, but this has the drawback that controlling NV-sample standoff is more difficult. One might expect that dropping the diamond sensor on a planar sample of interest would result in minimal standoff, but in practice the sample surface is usually not perfectly uniform or free of debris, which results in the diamond sensor sitting at an orientation which does not minimize standoff across the region of interest except through trial and error, as exemplified in Fig.~\ref{fig:spatial_res}(b). 

A possible solution is to mount the diamond sensor onto a stage and treat it as a ``probe'', in analogy with scanning probe microscopy techniques. By placing the sample on a separate positioning stage with angular degrees of freedom, it is then possible to control the orientation of the sample relative to the diamond sensor and align the two surfaces [Fig.~\ref{fig:easeofuse}(c)]. Together with a feedback mechanism, such as interference fringes, this strategy could allow consistent, repeatable standoff minimization, as was demonstrated in the case of relatively small planar probes (lateral size of order 10\,$\mu$m)  in Ref.~\onlinecite{ernstPlanarScanningProbe2019}. Extending this idea to larger diamond surfaces as required for widefield NV microscopy would be an important step towards streamlined interfacing.

In addition, as illustrated in Fig.~\ref{fig:easeofuse}, the widefield NV probe should integrate the microwave antenna [Fig.~\ref{fig:easeofuse}(b)], so that users only need to manipulate the sample stage for sample exchange. Similar to most scanning probe microscopies, the user should have the possibility to exchange the probe rapidly and effortlessly, for instance when a diamond with a special crystal orientation is needed, such as \{111\} or \{110\} orientations (see discussion in Sec.~\ref{subsec:sensor_design}). The probe could also be machined so that the surface area of the NV layer is reduced to its minimum required by the application. For instance, if a field of view of 100\,$\mu$m\,$\times$\,100\,$\mu$m is sufficient, a mesa of this size could be etched into the diamond surface, which could help minimize the standoff distance.

In the longer term, it should be possible to adapt this widefield NV probe concept for cryogenic measurements. One could even envision to make the probe interchangeable with commercial single-NV cantilevers and fit in existing standard cryostat systems. Such a plug-and-play integrated solution would greatly boost the appeal of the technique to potential users that are already equipped for scanning NV microscopy or magneto-optical microscopy at cryogenic temperatures.   

\subsubsection{Automation}

The controlled interfacing strategy described above is likely susceptible to automation. The diamond sensor and sample could be thus be brought together via an ``automated approach'', aligning the surfaces and minimizing standoff distance using control algorithms with interference fringes (or some other signal) as feedback. 

Many other tasks pertaining to the operation of the NV microscope can in principle be automated: auto-focusing the optics, aligning the bias field vector (e.g.\ via control of a Helmholtz coil system) to split the NV families, rotation of a half-wave plate in the excitation path to maximize ODMR contrast, and setting the microwave and pulsing parameters to optimize magnetic sensitivity. Additionally, measurement and processing of the ODMR signal, such as determining which microwave frequencies to scan and where the peaks are located could be tackled, for example, using Bayesian design.\cite{dushenkoSequentialBayesianExperiment2020} 

Such an automated setup would greatly improve the ease of use of the widefield NV microscope, and enable rapid sample exchange for high throughput measurements.

\subsubsection{Interchangeable optics}

For measurements where the resolution can be diffraction limited (rather than by NV-sample standoff or NV layer thickness) it may be desirable to use a low magnification, high field of view objective for initial alignment and holistic characterization or to find the area of interest, before switching to a high magnification and high numerical aperture objective for high resolution imaging. Therefore, easily interchangeable optics, e.g.\ through a selection of objective lenses mounted on a turret [Fig.~\ref{fig:easeofuse}(a)], would be a desirable feature. 

In this case, epifluorescence illumination is advantageous as the laser spot size will scale with the imaging field of view. Furthermore, the magnetic sensitivity per camera pixel is conveniently preserved upon changing objective, up to a correction factor to account for the change in collection efficiency, spin repumping efficiency and related effects. This conservation property is due to shot-noise sensitivity's equal and inverse scaling with image magnification and laser spot size diameter.\cite{glennMicrometerscaleMagneticImaging2017} One drawback of the epifluorescence geometry is that the sample is directly illuminated by the laser, which can cause significant heating. Alternatively, side illumination via a light sheet\cite{horsleyMicrowaveDeviceCharacterization2018} or total internal reflection\cite{lesageOpticalMagneticImaging2013} may be employed,  at the drawback of requiring another set of interchangeable optics to adjust the laser spot size.

\subsubsection{Rapid data processing}

In a typical experiment performed today, there can be a large time delay between signal acquisition and magnetic image data being available for sample analysis. In principle, the only physical limiting factor is the integration time, however in practice data transfer and processing also introduces a significant delay. In a widefield ODMR experiment, acquisition can easily reach data rates of order gigabits per second (Gbps). For example, a standard $4$\,megapixel camera with $30$\,ms exposure time and 16-bit depth produces data at $2$\,Gbps. While USB3 can handle this, processing such quantities of data as it arrives is nontrivial. For this reason, experiments generally proceed without continuous in-acquisition data processing. The ODMR data is then fit upon request by the user to quantitatively map parameters of interest, such as magnetic field. This processing may take up to several minutes, or even hours. 

However, modern data fitting techniques can dramatically speed up this processing time, for example via utilization of a Graphics Processing Unit (GPU) which can take advantage of the embarrassingly parallel problem of performing pixel-wise peak fitting.\cite{przybylskiGpufitOpensourceToolkit2017} Integrating and optimizing the data acquisition, processing and storage pipeline is feasible with present day technologies, and could enable the magnetic field map to be displayed and refreshed on a time scale of order seconds (i.e. after every single ODMR frequency sweep) [Fig.~\ref{fig:easeofuse}(d)]. Such rapid feedback would result in a user experience more akin to that of an optical microscope making NV microscopy more attractive for high throughput measurements.

\section{Conclusion}

In this Perspective we have outlined the proficiencies of widefield NV microscopy and the challenges that remain to its wider adoption as an analysis tool. Despite existing in a competitive market of microscopy techniques, we have identified particular niches such as the quantitative analysis of small amounts of magnetic materials embedded in a larger nonmagnetic
host, quantitative batch analysis of 2D magnetic materials, or millimeter-scale mapping of charge current distributions. Each of these applications have potential to be used by their specialist fields, as a primary technique or in complement of higher spatial resolution techniques, provided the microscope can be employed in a user-friendly manner. One major challenge to this end has been identified: the current lack of a simple repeatable sensor-sample interfacing method that does not hinder spatial resolution. 
Another improvement recognized is the need for automation and rapid data processing to increase throughput and ease of use. Continuing research programs include improvement of the NV-diamond sensor, further work to characterize the measurement accuracy, and benchmarking against other techniques in key applications.

With these developments, the widefield NV microscope may eventually become a commercially available tool that appeals to researchers from outside the NV community. Efforts are underway to create a general-purpose user-friendly instrument for room-temperature and low bias field operation. Such an instrument would be suitable for the characterization of many types of magnetic materials and current distributions. More advanced applications, such as for 2D magnetic materials, demand additional capabilities such as high bias fields and cryogenic operation. Here, there is an opportunity for developing a plug-and-play add-on fitting existing cryostat systems. The feasibility of such technological advances adds to the many exciting possibilities of NV sensing (such as the possibility of \textit{in situ} chemical analysis), confers good prospects for the future adoption of this technology and strongly motivates further work.

\begin{acknowledgments}
The authors acknowledge insightful discussions with David Simpson and Lloyd Hollenberg, and thank Cheng Tan and Lan Wang for the fabrication of the Fe$_3$GeTe$_2$ samples. J.-P.T. acknowledges support from the Australian Research Council (ARC) through grant FT200100073. S.C.S gratefully acknowledges the support of an Ernst and Grace Matthaei scholarship. A.J.H. is supported by an Australian Government Research Training Program Scholarship.

\end{acknowledgments}

\section*{Data Availability Statement}

The data that support the findings of this study, where not previously published, are available from the corresponding author upon reasonable request.

\bibliography{perspective}

\clearpage

\end{document}